\documentclass[prd,twocolumn,superscriptaddress,nofootinbib,10.5pt]{revtex4-1} 
\usepackage{XCharter}
\usepackage{mathptmx}
\usepackage[T1]{fontenc}

\usepackage{fullpage}
\usepackage{amsfonts}
\usepackage{amsmath}
\usepackage{slashed}
\usepackage{amssymb}
\usepackage{graphicx}
\usepackage{epic}
\usepackage{eepic}
\usepackage{epsfig}
\usepackage{pifont}
\usepackage{latexsym}
\usepackage[dvipsnames]{xcolor}
\usepackage[export]{adjustbox}
\usepackage{float}
\usepackage{multirow}
\usepackage[normalem]{ulem}
\usepackage{hyperref}
\usepackage{enumitem}
\usepackage{lipsum}
\usepackage[caption=false,labelformat=simple]{subfig}
 
\usepackage{natbib}
\usepackage{relsize}
\usepackage[left=1.6cm,right=1.6cm,top=1.58cm,bottom=1.58cm]{geometry}
\usepackage{comment}
\usepackage{dcolumn}
\usepackage{textcomp}
\usepackage{float}
\begin{document}
\title{Rescuing leptogenesis parameter space of inverse seesaw }
\author{Ananya Mukherjee}
\email{ananyatezpur@gmail.com}
\affiliation{Theory Division, Saha Institute of Nuclear Physics, 1/AF Bidhannagar, Kolkata
700 064, India} 
\author{Abhijit Kumar Saha}
\email{abhijit.saha@iopb.res.in}
\affiliation{Institute of Physics, Bhubaneswar, Sachivalaya Marg, Sainik School, Odisha 751005, India}
\begin{abstract}
{\noindent In a pure inverse seesaw framework, achieving a substantial lepton asymmetry that can be converted into the observed baryon asymmetry of the Universe is extremely challenging. The difficulty arises primarily due to two reasons, (i) partial cancellation of the lepton asymmetries associated with the components of a pseudo-Dirac pair, and (ii) strong wash out caused by the inverse decays. In this work we offer two possible resolutions to overcome the above mentioned challenges considering a (3,3) ISS framework. Our first proposal is based on the assumption of a non-standard cosmological era in the pre-BBN epoch, that triggers a faster expansion of the Universe, thereby reducing the washout by several orders of magnitude. The second proposition is an alternative of first which considers a quasi-degenerate mass spectrum for the singlet heavy neutrinos, resulting into a larger order of lepton asymmetry that survives the impact of strong washout to account for the observed BAU. The viable parameters space, as obtained can be tested at present and future Lepton Flavour Violation experiments {\it e.g.} MEG and MEG II.}
\end{abstract}
\maketitle

\section{Introduction}
 Seesaw mechanisms often provide a natural realisation of generating Standard Model (SM) neutrino masses and can also explain the origin of baryon asymmetry of the Universe ($\eta_B=(6.04-6.02)\times 10^{-10}$ \cite{ade2016planck}) via leptogenesis. In type I seesaw, a lepton asymmetry is created as a result of the lepton number and CP-violating out-of-equilibrium decay of the heavy right handed neutrinos (RHNs) \cite{Fukugita:1986hr,Buchmuller:2004nz} which subsequently gets partially converted to baryon asymmetry through the $(B + L)$ violating sphaleron processes. On the other hand, the inverse seesaw mechanism (ISS) mechanism \cite{Deppisch:2004fa} having a salient feature of offering the tiny neutrino mass at the cost of having a TeV scale heavy sterile states, makes a way to get itself verified in the collider experiments. Additionally, from theoretical perspective the ISS mechanism renders a large Yukawa coupling for a smaller mass scale of the heavy sterile neutrinos \cite{Dias:2012xp} which is in contrast to minimal type-I seesaw. Due to the TeV nature of these heavy sterile neutrinos present in the (3,3) ISS \footnote{\color{black} (2,2) and (2,3) schemes of ISS are also possible, predicting vanishing lightest SM neutrino mass \cite{Abada:2014vea}.}, we are allowed to investigate the flavor effects in leptogenesis \cite{Blanchet:2006be}.

In a TeV scale seesaw (violating the Davidson-Ibarra bound \cite{Davidson:2002qv}), one has to resort to resonant leptogenesis mechanism\,\cite{Blanchet:2006be,Deppisch:2010fr} in generating sufficient amount of asymmetry within the lepton sector that gets converted to the observed amount of baryon asymmetry via non-perturbative sphaleron processes before electroweak phase transition. {\color{black}Note that, we are especially interested in the scenario of producing lepton (baryon) asymmetry from heavy sterile neutrino decay in the TeV scale seesaw. Other mechanisms {\it e.g.} heavy neutrino oscillations, that offer a viable scenario of leptogenesis in ISS can be found in \cite{Abada:2015rta,Abada:2017ieq}.} Leptogenesis in pure ISS\footnote{We use the term ``pure" to highlight here the fact that, no other seesaw model has been considered in addition to the ISS.} with democratic or {hierarchical} structure of heavy sterile mass matrix 
encounters two serious challenges:\,(i) partial cancellation of lepton asymmetries among the pseudo Dirac states and (ii) huge wash out, predominantly induced by inverse decays. The partial cancellation of lepton asymmetries between pseudo-Dirac (PD) pairs kills the efficacy of resonant enhancement of individual flavor asymmetries and along with that, the impact of huge washout always make $\eta_B$ a few order smaller than its desired value \cite{ade2016planck}. 
{In fact, the work in \cite{Dolan:2018qpy} explicitly states that despite considering resonance effects, a pure inverse-seesaw scenario with either degenerate or hierarchical heavy sterile neutrino mass spectrum, fails to generate the required lepton (baryon) asymmetry.} Motivated by this, in the present article we propose two potential pathways to rescue ISS parameter space leading to a successful leptogenesis (baryogenesis) in the pure (3,3) ISS set up.  

An useful remedy to suppress the strong wash out of the produced lepton asymmetry is to make the Universe expand faster than the one with conventional radiation domination. This is possible if we assume the early Universe to be dominated by some son-standard fluid $\eta$ having equation of state parameter $\omega$ larger than of radiation \cite{DEramo:2017gpl}. Such assumptions are legitimate since the dynamics of pre-BBN Universe is vastly unknown which cannot be directly tested through experiments. In general, the energy density of $\eta$ is parametrised as $\rho_\eta\propto a^{-(4+n)}$ with $n>0$. The case of $n=2$ is familiar as kinaton domination or kination which is motivated by various cosmological events in the early Universe \cite{Gouttenoire:2021jhk}. For $n>2$, the early Universe encounters a faster than kination epoch.
Earlier works on leptogenesis \cite{Chang:2021ose,Chen:2019etb,Konar:2020vuu,Chakraborty:2022gob} in such specific kind of non-standard cosmological background have reported significant drop in the strength of washout and subsequently enhancement of relevant parameter space to a large extent, suitable for addressing baryogenesis. In the first part of our work, we intend to examine the validity of these non-standard cosmological models in order to rescue the ISS parameter space for leptogenesis. In another attempt, we consider a deviation from the democratic structure of the heavy sterile singlet neutrino {mass matrix}, which simply assumes a quasi degeneracy among {its elements}. This choice reduces the severity of partial cancellation of the lepton asymmetries between the pseudo-Dirac pairs and can successfully lead to adequate amount of baryon asymmetry, surviving the impact of strong washout. The two proposed solutions in order to restore the ISS parameter space for leptogenesis are independent and infact the second approach does not require the involvement of non-standard cosmology. Finally we discuss the testability of the obtained parameter spaces corresponding to each cases in upcoming high precision experiments on lepton flavour violation (LFV) searching for $\mu\to e\gamma$ decay \cite{Cheng:1980tp,BaBar:2009hkt,MEG:2013oxv,MEG:2016leq} and neutrino less double beta decay \cite{Andreotti:2010vj,CUORE:2015hsf,Klapdor-Kleingrothaus:2000eir,GERDA:2013vls,EXO-200:2014ofj,KamLAND-Zen:2012mmx,Bilenky:2014uka,Fang:2021jfv}. 

\section{Inverse seesaw}\label{sec:CI}
The extended type-1 seesaw mechanism, whose new physics can be essentially manifest at the TeV scale is familiar as the ISS mechanism proposed in Ref.~\cite{PhysRevLett.56.561,PhysRevD.34.1642}. The key feature of this seesaw realization comes through the soft explicit violation of lepton number which assists in bringing down the new physics scale to TeV without unnatural tuning of associated neutrino Yukawa couplings.
The ISS mechanism offers a sub-eV ordered neutrino mass at the cost of trading a set of SM gauge singlet (denoting as $S$) fermions along with the traditional three copies of TeV scale RHN ($N_{R}$). The following Lagrangian describes such mass generation mechanism of SM neutrinos under the ISS mechanism. 
\begin{equation}\label{ISS_Lag}
-\mathcal{L} \supset \lambda_{\nu}^{ \ell i} \,\overline{L_\ell} \,\widetilde{H} \,N_{R_i} + M_{R_{ij}} \,\overline{N_{R_i}^{C}} \,S_j + \frac{1}{2} \mu_{ij}\, \overline{S^C_i} S_j + h.c.
\end{equation}
with, $\widetilde{H} = i \sigma_2 H^*$. The $\ell,i$ being respectively the flavor and generation indices of leptons, RHNs and newly added SM gauge singlet fermions. The complete mass matrix constructed from the basis $( \overline{\nu_L}, \overline{N_R^C}, \overline{S^C})$ with the help of Eq.~\ref{ISS_Lag} can be written in the following texture:
\begin{equation}
M_\nu = \begin{pmatrix}
0 && m_D && 0 \\
m_D^T && 0 && M_R \\
0 && M_R^T && \mu \\
\end{pmatrix},\label{eq:totalnuM}
\end{equation}  
where, $m_D,\, M_R\,\, \text{and}\,\, \mu$ are $3\times3$ mass matrices. We have considered here $M_R$ and $\mu$ to be diagonal having the structures $M_R={\rm diag}\{M_1,M_2,M_3\}$ and $\mu={\rm diag}\{\mu_1,\mu_2,\mu_3\}$, indicating minimal lepton number violation. A democratic structure of ISS implies $M_1=M_2=M_3$ and $\mu_1=\mu_2=\mu_3$. The light neutrino mass can be found by performing a block diagonalization of $M_\nu$ matrix, which leads to the following,
\begin{equation}\label{eq:mnu}
m_\nu = m_D (M_R^T)^{-1}\mu M_R^{-1}m_D^T.
\end{equation}
The $6\times6$ heavy neutrino mass matrix in the basis $(N_R, S)$ can be put into the following form,
\begin{equation}
M_N^{6\times6} = \begin{pmatrix}\label{6by6}
0 && M_R \\
M_R^T && \mu  \\
\end{pmatrix},
\end{equation}
For a single generation of $N_R$ and $S$, the $M_N$ would be a $2\times 2$ matrix. In that case, one can write the resulting pseudo-Dirac mass states having the following form~\cite{Dolan:2018qpy}, 
\begin{equation}\label{eq:splitting}
M_N =  \frac{1}{2} \left( \mu \pm \sqrt{\mu^2 + 4M_R^2}\right),
\end{equation} 
with $\mu$ as the same lepton number violating scale which essentially acts as the source of tiny non-degeneracy among the final pseudo-Dirac pairs in the ISS model. 

The extraction of the ISS Yukawa coupling through the CI parametrization is obtained as \cite{Dolan:2018qpy},
\begin{equation}
\lambda_{\nu} = \frac{1}{v} \,U_{\rm PMNS} \,m_{n}^{1/2} \,\mathbb{R} \, \mu^{-1/2}\,M_{R}^T\,,
\end{equation}
where $\,m_n$ is the $3\times3$ mass matrices having definitions $m_n \equiv \text{diag}(m_{1},m_{2},m_{3})$. Without loss of generality, we assume $\mu$ also to be diagonal. Here $U_{\rm PMNS}$ is the standard PMNS matrix and $v=174$ GeV represents the SM Higgs vacuum expectation value (VEV). In general, $R$ is a complex orthogonal matrix {\it i.e.} $\mathbb{R} \mathbb{R}^{T}={\bf \mathbb{I}}$. However, one can choose a real $\mathbb{R}$ in order to proceed with minimal number of CP violating sources. Here we have worked with the conventional form of $\mathbb{R}$ as given by,
\begin{align}\label{eq:CIRot}
\mathbb{R}=\left(
\begin{array}{ccc}
 c_y c_z & -s_ x s_y c_z- c_x s_z & s_x s_z- c_x s_y c_z \\
 c_y s_z & c_x c_z-s_x s_y s_z & -s_x c_z-c_x s_y s_z \\
 s_y & s_x c_y & c_x c_y \\
\end{array}
\right),
\end{align}
 where $c,\,s$ represent cosine and sine of the rotational angles $x,y$ and $z$ , which can be both real or complex. Considering the normal hierarchical scenario, we write the mass eigenvalues of SM active neutrinos as $m_{2}=\sqrt{m_l^2+\Delta m_{\rm sol}^2}$ and $m_3=\sqrt{m_2^2+\Delta m_{\rm atm}^2}$ where we have used the best fit values for $\Delta m_{21}^2$
 and $\Delta m_{32}^2$\,\cite{ParticleDataGroup:2022pth} with $m_l$, stands for the lightest active neutrino mass. We have also set the neutrino mixing angles at their corresponding central values\,\cite{ParticleDataGroup:2022pth}. In the $U_{\rm PMNS}$ we have set the neutrino mixing angles at their central values as well\,\footnote{For the case A we fix Dirac phase (appearing in $U_{\rm PMNS}$) as $\delta_{\rm CP}=3\pi/2$, whereas we choose a varying $\delta_{\rm CP}$ from $\pi/6\,-\,3\pi/2$ for the case B.}.

\subsection{Deviation from unitarity of the $U_{\text{PMNS}}$} \label{NU}
In general, non-unitarity (NU) appears whenever additional heavy sterile states mix with the light SM neutrinos similar to the current scenario. Note that, the diagonalization of $m_\nu$ by $U_{\rm PMNS}$ does not diagonalize the $M_R$ and $\mu$. 
One can express the diagonalising matrix of $M_\nu^{9 \times 9}$ as, 
\begin{equation}\label{eq:NUmain}
V = \left( \begin{array}{cc}V_{3 \times 3} &V_{3 \times 6} \\
V_{6 \times 3} & V_{6 \times 6}\end{array}\right)
\end{equation}
with, $M_\nu^{\rm diag} = V^T M_{\nu} V$.
The $V$ matrix in the above equation takes the following conventional form \cite{Korner:1992zk,Das:2014jxa,Dev:2012sg},
\small{\begin{align}
V = \left( \begin{array}{cc}  (1_{3\times3} + \zeta^* \zeta^T)^{-1/2} & \zeta^*(1_{6\times6} + \zeta^T\zeta^*)^{-1/2}\\
-\zeta^T(1_{3\times3} + \zeta^*\zeta^T)^{-1/2} & (1_{6\times6} + \zeta^T \zeta^*)^{-1/2}\end{array}\right)  \left(\begin{array}{cc}U_{\rm PMNS} & 0\\0& V^\prime \end{array}\right)
\end{align}}
 where $V^\prime_{6\times 6}$ is the unitary matrix that diagonalise the heavy neutrino submatrix in $M_\nu$ and $\zeta = (0 _{3\times3}, m_D M_R^{-1})$ is a ${3\times 6}$ matrix. With the assumption of a minimal flavor violation implying $M_R$ and $\mu$ to be diagonal, $V^\prime$ can be evaluated as,
\begin{equation}
V^{\prime} = \frac{1}{\sqrt{2}} \left(\begin{array}{cc} 1_{3 \times3} & -i 1_{3 \times3} \\ 
1_{3 \times 3}& i 1_{3 \times 3}\end{array} \right)+ \mathcal{O} (\mu M_R^{-1})
\end{equation}
Next we derive the analytical form of each components of $V$ in Eq.(\ref{eq:NUmain}). The sub-block $V_{3\times 3}$ is found to be,
\begin{equation}\label{eq:NUsub1}
V_{3\times 3} = \Big(1_{3\times3} + \zeta^* \zeta^T\Big)^{-1/2} U_{\rm PMNS}  \simeq (1_{3\times3} - \eta) U_{\rm PMNS}
\end{equation}
in the $|\zeta|\ll 1$ limit. The $V_{3\times 3}$ is the matrix that includes the correction of NU over the original PMNS matrix that exactly diagonalises $m_\nu$.
The quantity $\eta \equiv  1_{3\times3}-(1_{3\times3} + \zeta^* \zeta^T)^{-1/2}$ in Eq.(\ref{eq:NUsub1}) measures the deviation from Unitarity. Experimental bound on the elements of $\eta$ can be found from \cite{Fernandez-Martinez:2016lgt} as given below.
\begin{equation}
 |\eta_{i j}| \leq \left(
\begin{array}{ccc}
1.3 \cdot 10^{-3}  & 1.2 \cdot 10^{-5} & 1.4 \cdot 10^{-3} \\ 
1.2 \cdot 10^{-5}  & 2.2 \cdot 10^{-4} & 6.0 \cdot 10^{-4} \\ 
1.4 \cdot 10^{-3}  & 6.0 \cdot 10^{-4} & 2.8 \cdot 10^{-3} 
\end{array}
\right) .
\label{eq:NU}
\end{equation} 
Thus the emergence of NU in the present set up holds the potential to directly serve as a viable tool of testing the relevant model parameter space. Also, the non-unitarity matrix elements may leave non-trivial roles in the branching of LFV decays.

\section{Leptogenesis}\label{sec:lep}
In the ISS model the decay of the pseudo-Dirac neutral states trigger the generation of lepton asymmetry. These pseudo-Dirac neutral states (denoted by $N_k$) undergo CP violating decay into the SM lepton ($L_{\ell}$) and the Higgs doublet ($H$) as, 
    \begin{equation}
    \epsilon_{N_{k}}^\ell = -\sum \frac{\Gamma(N_{k} \rightarrow L_{\ell}+H^{+} , \nu_l+ H^0)-\Gamma(N_{k} \rightarrow L_{\ell}+H^-, \nu_\ell^c +H^{0^*})}
    {\Gamma(N_{k} \rightarrow L_{\ell}+H^{+} , \nu_\ell +H^0)+\Gamma(N_{k} \rightarrow L_{\ell}+H^{-}, \nu_\ell^c +H^{0^*})}
   \end{equation} 
As evident from the above expression, the CP-asymmetry is a measure of the difference in decay widths of $N_k$ to a process and its conjugate process. At tree level, these two are the same resulting into vanishing CP-asymmetry. Taking into account the one loop vertex and self energy diagrams, it is found that non-zero CP-asymmetry arises due to the interference between the tree level and the one loop diagrams. For the decaying pseudo-Dirac mass falling in the TeV regime, lepton asymmetry gets maximum contribution from the self-energy diagram \cite{Pilaftsis:2005rv}. In such scenario, the individual lepton asymmetries are resonantly enhanced and can reach even $\mathcal{O}(1)$. 

 In order to compute the $\epsilon$ and $\eta_B$ from the decay of pseudo-Dirac eigenstates, one must pursue change-of-basis exercise for the Yukawa couplings such that heavy neutrino mass matrix ($M_N^{6\times 6}$ in Eq.(\ref{eq:totalnuM})) is diagonal real and positive. We write the modified Yukawa couplings as $Y_{\nu_{ij}}$, appearing in the the complete structure for $9\times 9$ neutrino mass matrix after the sub-block diagonalisation of $M_\nu$ in Eq.(\ref{eq:totalnuM}),
 \begin{align}
     M_{\nu}^\prime=\begin{pmatrix}
     0 & m_D^\prime & 0\\
     m_D^{\prime T} & M_{E_1} & 0\\
     0 & 0 &M_{E_2}\\
     \end{pmatrix}\,,
 \end{align}
 where $m_D^\prime=\frac{Y_\nu v}{\sqrt{2}}$. The $M_{E_1}$ and $M_{E_2}$ are two $3\times 3$ diagonal matrices with the elements representing the eigenvalues of the physical eigenstates and their pseudo Dirac partners respectively.
As mentioned earlier, flavor-dependent effects of leptogenesis are relevant at low enough temperatures (set by the RHN mass) such that at least one charged leptons is in thermal equilibrium. When this condition is met, flavor-dependent effects are not avoidable as the efficiency factors differ significantly for the distinguishable flavors. We use the following standard expression for the lepton asymmetry parameter \cite{Adhikary:2014qba} \footnote{\color{black}In our work we do not consider the finite temperature effect in the computation of CP asymmetry parameter. In appendix \ref{sec:FTbaryonass}, we have shown that temperature correction leaves very minimal impact on the final baryon asymmetry, even in case of resonant leptogenesis.},
\begin{widetext}
\begin{align}\label{eq:asymmetry}
    \epsilon_{i}^{\ell} = & \frac{1}{8\pi \left(Y_\nu^{\dagger} Y_\nu\right)_{ii}}\sum_{j \neq i} \text{Im}\left[\left(Y_{\nu}^\dagger Y_{\nu}\right)_{ij}\left(Y_\nu^{\dagger}\right)_{i\ell}\left(Y_{\nu}\right)_{\ell j}\right]  \left[ f(x_{ij})+ \frac{\sqrt{x_{ij}} \left(1 - x_{ij}\right)}{\left(1- x_{ij}\right)^2 + \frac{1}{64 \pi^2} \left(Y_{\nu}^\dagger Y_{\nu}\right)_{jj}^2}\right]  + \frac{1}{8\pi \left(Y_\nu^\dagger Y_\nu\right)_{ii}}\sum_{j \neq i} \frac{ (1- x_{ij}) \text{Im}\left[\left(Y_{\nu}^\dagger Y_{\nu}\right)_{ij}\left(Y_\nu^{\dagger}\right)_{i\ell}\left(Y_{\nu}\right)_{\ell j}\right] }{\left(1- x_{ij}\right)^2 + \frac{1}{64 \pi^2} \left(Y_{\nu}^\dagger Y_{\nu}\right)_{jj}^2},
   \end{align}
   \end{widetext}
with the following definition for the loop function 
$ f(x_{ij}) = \sqrt{x_{ij}} \left[1 - (1+x_{ij})\text{ln}\left(\frac{1-x_{ij}}{x_{ij}}\right) \right]$ 
where, $x_{ij} = \left(\frac{M_{N_j}}{M_{N_i}}\right)^2$.

The set of Boltzmann equations that governs the dynamics of decay of the heavy pseudo-Dirac states and yield of lepton asymmetries are the following \cite{Pilaftsis:2005rv}:
\begin{align}
\label{beq1} 
\frac{d \eta_{N_{i}}}{dz} &= \frac{z}{\mathcal{H}(z=1)}\
\Bigg[\,\Bigg( 1 \: -\: \frac{\eta_{N_{i}}}{\eta^{\rm eq}_{N_{i}}}\,
\Bigg)\, \sum_{k\,=\,e,\mu,\tau}
\Gamma^{D\; (i k)} -
\frac{2}{3}\, \sum_{k\,=\,e,\mu,\tau} \eta^{k}_{\ell}\,
\varepsilon^{k}_{i}\,\nonumber\\
&~~~~~~~~~~~~~~~~~~~~~~~~~~~~~~\times 
\widehat{\Gamma}^{D\; (i k)}\,\Bigg]\,,
\end{align}
\begin{align}
  \label{beq2} 
\frac{d \eta_\ell}{dz} &= 
\frac{z}{\mathcal{H}(z=1)}\, \Bigg[\, \sum\limits_{i=1}^2\,
\varepsilon^{\ell}_{i}\ \Bigg(
\frac{\eta_{N_{i}}}{\eta^{\rm eq}_{N_{i}}} \: -\: 1\,\Bigg)\, 
\sum_{\beta\,=\,e,\mu,\tau} 
\Gamma^{D\; (i k)} \nonumber\\ 
&~~~~~~~~~~~~~~~~~~~~~~-\,\frac{2}{3}\, \eta_\ell,  \sum\limits_{i=1}^2\,
B^{\ell}_{i}\, \widetilde{\Gamma}^{D\; (i \ell)} \:\Bigg]\,,
\end{align}
where we have defined $z=\frac{M_{N_1}}{T}$. The quantities $\eta_{N_i}$ and $\eta_l$ are the number densities of $i^{\rm th}$ pseudo-Dirac state, and created lepton asymmetry normalised to the photon number density, such that one can write $\eta_a(z) = n_a(z)/n_\gamma (z)$ with $n_\gamma (z) = 2m_{N_1}^3/\pi^2\times 1/z^3$. The R.H.S of Eqs.~\eqref{beq1} and \eqref{beq2} in principle should involve various $2\to 2$ processes in addition to $1\to 2$ processes, which are calculated in Ref.~\cite{Pilaftsis:2005rv}. We have chosen to neglect the contributions from all $2\to 2$ scattering processes since their impact on erasing the generated lepton asymmetry is negligible compared to the inverse decays \cite{Buchmuller:2004nz,Pilaftsis:2005rv,Deka:2021koh,Dey:2021ecr}. The Hubble parameter of the Universe is represented by $\mathcal{H}$ in Eq.(\ref{beq1}).
The total decay width $\Gamma_{N_i}$ of the RHNs is given by
\begin{equation}
\Gamma_{N_i}\ =\ \sum_{l=1}^{3} \Gamma^{\ell}_{N_i} =\ \frac{M_{N_i}}{8\pi}\
\sum_{l=1}^{3} Y_{\nu_{i\ell}}^{\dagger} Y_{\nu_{i\ell}} .
\end{equation}
The rate for a generic process $X\to Y$ and its conjugate counterpart $\overline{X}\to \overline{Y}$ is defined as $\gamma^X_Y$.
For   the  $1\to   2$   process,\   $N_i\to   L\Phi$  or   $N_i\to
L^C\Phi^\dagger$, 
$\gamma^{X}_{Y}$ is given by \cite{Pilaftsis:2005rv}
\begin{eqnarray}
\gamma^{N_i}_{L_\ell \Phi}\ = \frac{M_{N_1} M_{N_i}^{2} \Gamma_{N_{i}}^{\ell} }{\pi^2\, z}\ K_1(z \sqrt{a_i})\,,
\end{eqnarray}
in terms of the re-scaled variables of Eq.~\eqref{rescaled} where $K_n(z)$ is an $n^{\rm th}$ order modified Bessel function. 
\begin{equation}\
\label{rescaled}
z = \frac{M_{N_1}}{T}\,,  \;\; 
a_i = \left(\frac{M_{N_i}}{M_{N_1}}\right)^2\,, \;\;  
\end{equation}
with  $s$ being the Mandelstam variable \cite{Luty:1992un}. The following definitions are implemented in the aforementioned coupled BEQs. For a detailed description one may look into \cite{Pilaftsis:2005rv}.
\begin{align}\label{eq:washoutE}
\Gamma^{D(i \ell)} & =  \frac{1}{n_\gamma}\
\gamma^{N_{i}}_{L_\ell\Phi}\nonumber\\
\widehat{\Gamma}^{D(i \ell)} & = 
\frac{1}{n_\gamma}\left(1+\frac{4}{21}\,\frac{\eta_{\Delta L}}{\eta_{\Delta L_\ell}} \right)\,
\gamma^{N_{i}}_{L_\ell\Phi}\nonumber\\
\widetilde{\Gamma}^{D(i \ell)} & = 
\frac{1}{n_\gamma}\left(1+\frac{4}{21}\,\frac{\eta_{\Delta L}}{\eta_{\Delta L_\ell}} \right)\,
\gamma^{N_{i}}_{L_\ell\Phi}
\end{align}

In a standard radiation dominated Universe, the analytically approximated solution of baryon to photon ratio is given by \cite{Pilaftsis:2003gt,Deppisch:2010fr,Bambhaniya:2016rbb},
 \begin{equation}\label{Eq:bau}
 \eta_B \simeq -3\times 10^{-2} \sum_{\ell,i}\frac{\epsilon_{i \ell}}{K_\ell^{\text{eff}}\text{min}\left[z_c,1.25 \, \text{Log}\,(25 K_\ell^{\text{eff}})\right]} ~,
\end{equation}  
where $z_c = \frac{M_i}{T_c}$ and $T_c \sim 149$ GeV, \cite{Bambhaniya:2016rbb} is the critical temperature, below which the sphalerons freeze out \cite{PhysRevD.49.6394,DOnofrio:2012phz}. Here,  $K_\ell^{\text{eff}} = \kappa_\ell \sum_{i} K_i B_{i\ell}$, with $K_i = \Gamma_{N_i} / \zeta(3)\mathcal{H}(z=1)$, denoting the washout factor. Here, $B_{i\ell}$'s are the branching ratios of the $N_i$ decay to leptons of $\ell^{th}$ flavor : $B_{i\ell} = \frac{|Y_{\nu_{i\ell}}|^2}{(Y_{\nu}Y_{\nu}^{\dagger})_{ii}}$.
Including the Real Intermediate State~(RIS) subtracted collision terms one can write the factor $\kappa_\ell$ as,
\begin{align}
\kappa_\ell = & 2 \sum_{i,j j \neq i} \frac{\text{Re}\left[(Y_{\nu})_{i\ell}(Y_{\nu})_{ j\ell}^* \left(Y_{\nu} Y_{\nu}^\dagger\right)_{ij}\right]+ \text{Im}\left[\left(\left(Y_{\nu}\right)_{ i\ell} (Y_{\nu})_{ j\ell}^*\right)^2\right]}{\text{Re}[(Y_{\nu}^\dagger Y_{\nu})_{\ell \ell} \{(Y_{\nu} Y_{\nu}^\dagger)_{ii} + \left(Y_{\nu}Y_{\nu}^\dagger\right)_{jj}\}]} \nonumber\\
& ~~~~~~~~~~~~~~~~~~~~~~~~~~~~~~~~~~~~~~~~\times\left(1-2i \frac{M_i-M_j}{\Gamma_i + \Gamma_j}\right)^{-1}.
\end{align}

Leptogenesis in the ISS model suffers from it's incapacity to explain the observed BAU primarily due to two reasons. One is the impact of huge washout and the second one is inadequate amount of flavor lepton asymmetry, generated from a pseudo-Dirac pair (see e.g. \cite{Dolan:2018qpy}). Authors, in \cite{Dolan:2018qpy} show that even after encountering resonance it is impossible to obtain adequate amount of lepton asymmetry required for achieving the observed BAU. As mentioned in the introduction, here we particularly deal with these two challenges in order to evade the competition between lepton asymmetry and washout, by two different approaches.

The first approach we adopt here, in order to make the ISS framework viable for leptogenesis involves the modification of the Hubble expansion rate during lepton asymmetry creation and subsequent washout process. The washout factor is mainly determined by comparing the decay rate with the Hubble expansion rate of the Universe at a certain temperature T as defined by the parameter $K_i$ earlier. The Hubble expansion rate in a radiation dominated universe fails to compete against the larger decay width of the heavier pseudo-Dirac states of the ISS framework and hence a modification of the standard description of cosmology resulting into a faster expansion of the Universe could be useful in this regard.
In the second approach, the partial cancellation of lepton asymmetries between pseudo Dirac pairs can be undone by considering non-degenerate diagonal entries of $M_R$ matrix. For analytic understanding one has to look into the Yukawa texture of the ISS model which determines the overall order of lepton asymmetry contributed by the decay of the final pseudo-Dirac states.  
\section{Constraints from $\mu \rightarrow e  \gamma$}\label{sec:lfv}
In this work, we have focused only on the study of the particular branching ratio (BR) for $ \mu \rightarrow e  \gamma $ decay process, which presently provides the strongest bound in comparison to other variants of LFV decay. In the ISS scenario, one can naturally obtain a large branching of these LFV decays in comparison to what one obtains in the type-I seesaw mechanism\footnote{As we have seen, the rates of the LFV processes in the canonical type-I seesaw model with massive right handed neutrinos are so strongly suppressed that these processes are not observable in practice, and one has {\it e.g.} BR$(\mu \rightarrow e  \gamma ) \lesssim 10^{-47}$ \cite{Cheng:1980tp,Aubert:2009ag} in a high scale type-I seesaw scenario.}. This large BR (here in particular, BR$(\mu \rightarrow e \gamma)$) is in practice resulted from the large light-heavy mixing (denoted by $V_{\mu i}, \,\,\rm{and}\,\,V_{e i} $) mentioned in the Eq.(\ref{eq:light-heavy}). The conventional form of BR$(\mu \rightarrow e \gamma)$) is given by \cite{Abada:2014vea,Korner:1992zk},
\begin{equation}\label{eq:light-heavy}
\text{BR} (\mu \rightarrow e\gamma) = \frac{\alpha_w^3 s_w^2}{256 \pi^2}\frac{m_\mu^5}{M_W^4}\frac{1}{\Gamma_\mu} \Big| \sum_i^9 V^*_{\mu i} V_{e i}G(y_i) \Big|^2,
\end{equation}
where, $\alpha_w = g_w^2/ 4\pi$ and $s_w^2 = 1-(M_W/M_Z)^2$ along with the loop function $G(y)$ having the following form,
\begin{equation}
G(y) = - \frac{2 y^3 + 5 y^2 - y}{4(1-y)^3}- \frac{3y^3}{2(1-y)^4}\text{ln} y ,~~~~ \text{with}\,\, y_i =\frac{m_i^2}{M_W^2}.
\end{equation}
\vspace{0.5cm}

Here, $M_W$ and $M_Z$ imply the masses of the $W$ and $Z$ bosons that participate in the loop diagrams of the flavor violating decay of our interest. One denotes $\Gamma_\mu$ as the decay width of the relevant decay. In the above equation $m_i$ stands for both the active and all the sterile neutrino mass states and $V$ being the NU mixing matrix as defined earlier. We would like to refer the reader to Sec. \ref{NU} for the construction of such mixing matrix.

\section{Results and analysis}
In this section, we present our claims regarding the possibilities that we have explored in finding the viable parameter space for leptogenesis in ISS. As mentioned in the introduction, we have proposed two different kinds of resolutions to the issue of not having successful leptogenesis in a generic ISS framework. In particular, the two case studies are based on (i) reducing the washout effects by a considerable amount, and (ii) increasing the total lepton asymmetry contributed by all the pseudo-Dirac pairs of the ISS framework, to the possible maximum order ($\mathcal{O}(1)$). These two approaches involve two sub cases resulted from two different choices of the rotational matrix used to extract the Yukawa coupling in the CI formalism. One is considering a complex $\mathbb{R}$ and the other is with a real $\mathbb{R}$ matrix, both of which satisfy $\mathbb{R} \mathbb{R}^T \,=\, \mathbb{I}$. A complex $\mathbb{R}$ assists in raising the order of the Yukawa coupling ($Y_{\nu_{i\ell}}$) as it involves hyperbolic function of the rotational angles in the CI parametrisation. Such rise in the order of $Y_{\nu_{i\ell}}$ is not expected in case of the real $\mathbb{R}$. On the other hand a larger Yukawa generally leads to large values of light-heavy mixing, which in turn influences the order of magnitude of branching ratios of the LFV channels. As mentioned in Sec.~\ref{sec:lfv} an indirect probe of the leptogenesis parameter space is possible in the LFV experiments. 

\begin{center}
\begin{table*}
\begin{center}
  \renewcommand*{\arraystretch}{1.4}
\begin{tabular}{| c | c | c | c | c | c | c | c | c | c | c | c | c | c | c |}
  \hline
~BP\,-\,CA\, ~&$\mu$\,(GeV)& $m_l$\,(eV) &~~ $x$~~ &~~ $y$~~ & ~~$z$~~  & $~n~$ & $~T_r$\,(MeV)~ & $ \Gamma^{D,1}/\mathcal{H}$ ($z =1 $)  & $\eta_B$ & ~Br$(\mu\to e\gamma)$~ & ~$m_{\beta\beta}$\,(eV) ~ & ~NU~~\\
    \hline
      I & $~2.65\times 10^{-3}~$ & $~5.94\times 10^{-4}~$ &  ~0.63~ & ~3.02~ & ~3.05~ & ~2~ &  ~5~ & ~9.66~& ~$6.02 \times 10^{-10}$~ &~ $1.06\times 10^{-20}$~&~ 0.0019 ~&~\checkmark ~\\ 
  \hline
  II & $4\times 10^{-6}$ & $3\times 10^{-4}$  & 0.33 & 1.44 &  1.19 & 3 & 5 &   $0.16$   & $6.10\times10^{-10}$ & ~$8.50\times 10^{-15}$~& 0.0017 &~\checkmark ~ \\ 
  \hline
\end{tabular}
\caption{Benchmark choices for the relevant parameters of leptogenesis and the corresponding washout amount (with $\Gamma^{D,1}=\sum_{l=1}^3\Gamma^{D(l1)}$ in Eq.(\ref{eq:washoutE})) which altogether yield the final baryon to photon ratio ($\eta_B)$, considering a real $\mathbb{R}$. The $\eta_B$ values for both the benchmark points are evaluated at $z = 100$ by solving the set of coupled Boltzmann equations. Note that in this case the Dirac CP phase is the only parameter that leads to the CP violation. We also mention the estimates of Br$(\mu\to e\gamma)$ and $m_{\beta\beta}$ for the two respective benchmark points. Here we have set $M_1=M_2=M_3=1$\,TeV. The last column reveals whether a particular BP is allowed (\checkmark) or disallowed (\ding{55}) by the NU constraints.}
\label{tab:CIreal}
\end{center} 
\end{table*}
\end{center}

\begin{figure*}[htb!]
    \centering
    \includegraphics[height=4.4cm,width=4.6cm]{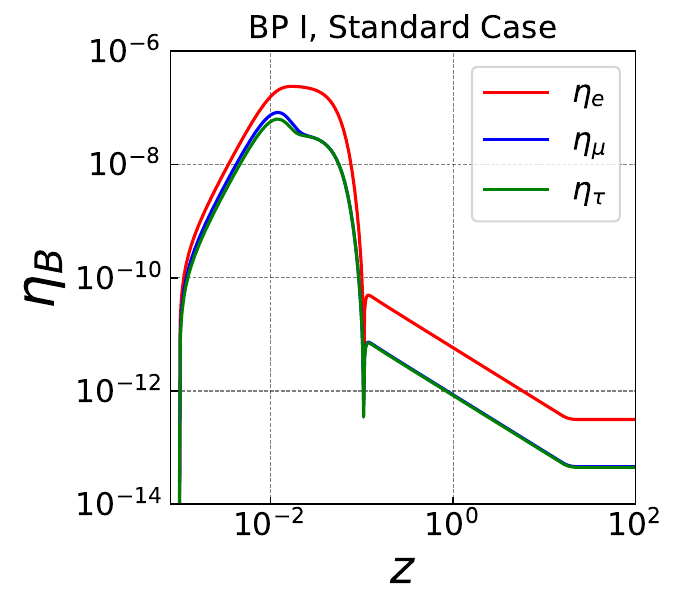}~
    \includegraphics[height=4.4cm,width=4.6cm]{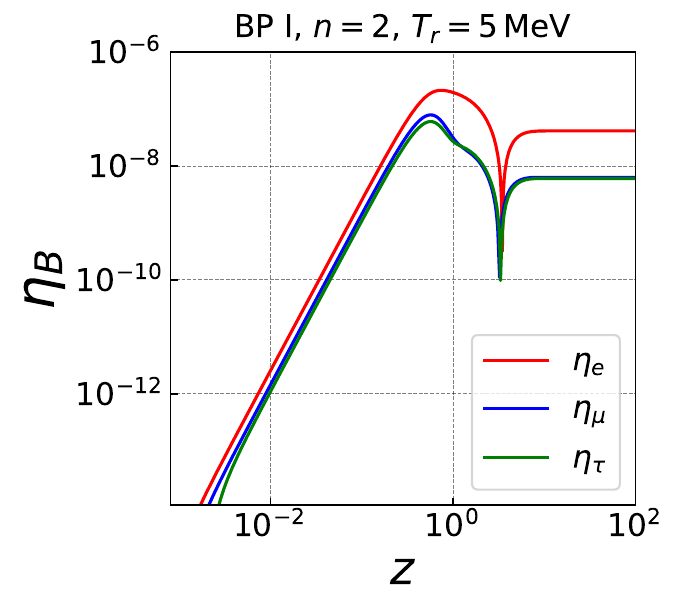}~
    \includegraphics[height=4.4cm,width=4.6cm]{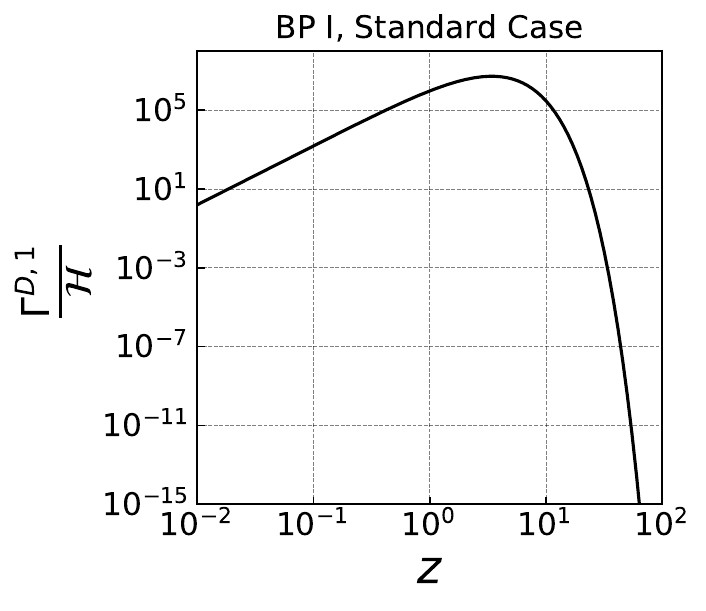}~
    \includegraphics[height=4.4cm,width=4.6cm]{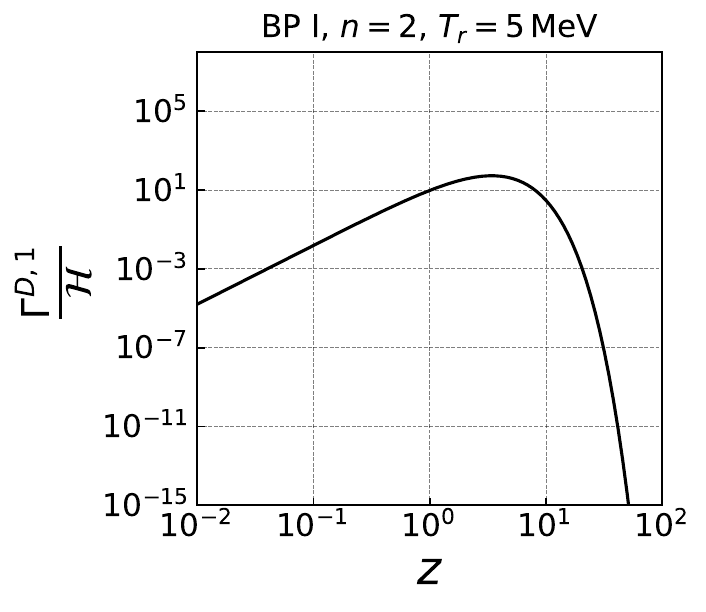}
    \caption{In the first two columns, the evolution of lepton asymmetry yield, $\eta_{i}$ is shown as a function of temperature for BP I considering standard RD Universe and nonstandard cosmological scenario respectively. The last two columns show the evolutions of the ratio $(\Gamma^{D,1}/\mathcal{H})$ with temperature for standard and nonstandard cosmology respectively.}
    \label{fig:C1BPIreal}
\end{figure*}
\begin{figure*}[htb!]
    \centering
    \includegraphics[height=4.4cm,width=4.6cm]{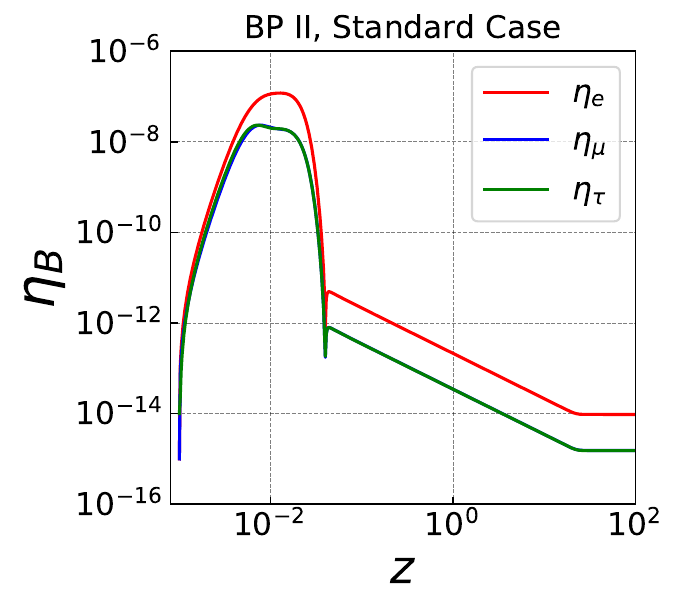}~
    \includegraphics[height=4.4cm,width=4.6cm]{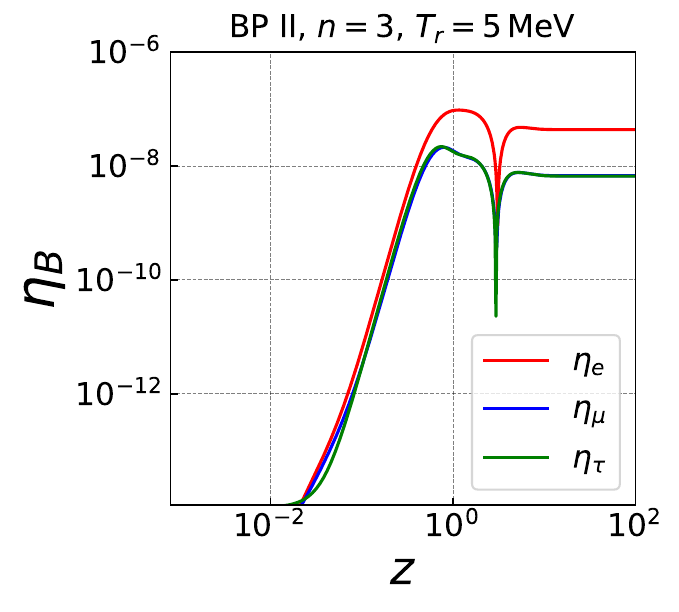}~
    \includegraphics[height=4.4cm,width=4.6cm]{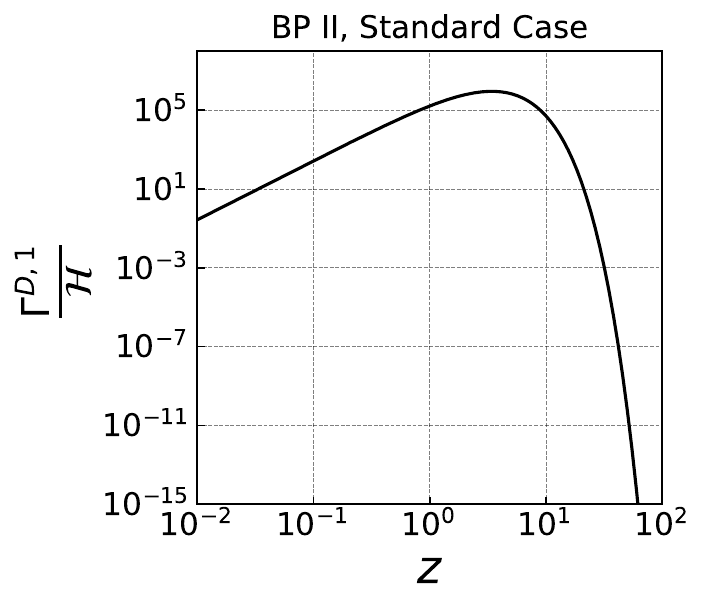}~
    \includegraphics[height=4.4cm,width=4.6cm]{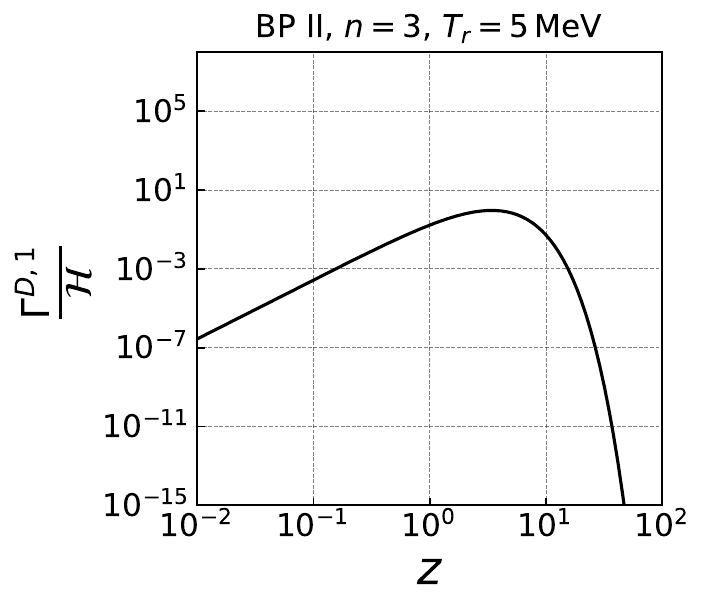}
    \caption{In the first two columns, the evolution of lepton asymmetry yield, $\eta_{i}$ is shown as a function of temperature for BP II considering standard RD Universe and nonstandard cosmological scenario respectively. The last two columns show the evolutions of the ratio $(\Gamma^{D,1}/\mathcal{H})$ with temperature for standard and nonstandard cosmology respectively.}
    \label{fig:C1BPIIreal}
\end{figure*}

\subsection{Case I: Reducing the strength of wash-out}\label{non-standard}

The Hubble parameter $\mathcal{H}$, being connected to the total energy density of the Universe through standard Friedmann equation sets the expansion rate of the Universe as function of temperature. In conventional standard cosmology, it is assumed that the Universe remains radiation dominated from the end of reheating epoch till BBN. Non-standard scenarios appear due to the presence of a non-trivial fluid that dominates the energy density of the Universe having equation of state parameter $\omega\neq \frac{1}{3}$ at an intermediate stage. For example, an early matter dominated epoch arises in the early Universe if the additional fluid has $\omega=0$ and exceeds the radiation energy density of the Universe at some point of time. Alternatively the pre-BBN era could be occupied by a species with $\omega>\frac{1}{3}$ besides the conventional radiation component. The later scenario is of our interest in the present analysis which is usually dubbed as {\it fast expanding Universe}.

 For convenience we mark the new species with $\eta$ and write the corresponding energy density as $\rho_\eta\sim a^{-4(1+n)}$, where $a(t)$ is the scale factor of the Universe. In the limit of entropy conservation per comoving volume \textit{i.e.,} $s\,a^3=$ const., one can define $\rho_{\text{rad}}(t)\propto a(t)^{-4} $. Now, in case of a faster expansion of the Universe the energy density of $\eta$ field is anticipated to be red-shifted more rapidly than the radiation as realized by $n>0$. Utilizing the energy conservation principle, a general form of $\rho_\eta$ can be constructed as:
\begin{equation}
 \rho_\eta(T) =  \rho_\eta(T_r)\,\left(\frac{g_{*s}(T)}{g_{*s}(T_r)}\right)^{(4+n)/3}\left(\frac{T}{T_r}\right)^{(4+n)}.
\end{equation}
\noindent The temperature $T_r$ is an unknown parameter ($ >T_{\text{BBN}}$) and can be safely assumed as the point of equality of two respective energy densities: $\rho_\eta(T_r)=\rho_{\text{rad}}(T_r)$. To keep the success of BBN intact, the energy component $\rho_\eta$ must be subdominant compared to $\rho_R$ before BBN takes place.  This poses a bound on $T_r$ as $T_r\gtrsim (15.4)^{1/n}$ \cite{DEramo:2017gpl}.

\begin{widetext}
\begin{center}
\begin{table}[h]
  \renewcommand*{\arraystretch}{1.4}
\addtolength{\tabcolsep}{-2.5pt}
\begin{center}
\begin{tabular}{| c | c | c | c | c | c | c | c | c | c | c | c | c | c | c |}
  \hline
~BP\,-\,CA ~&$\mu$\,(GeV)& $m_l$\,(eV) &~~ $x$~~ &~~ $y$~~ & ~~$z$~~  & ~$~n~$~ & $~T_r$\,(MeV)~ & ~$\frac{\Gamma^{D,1}}{\mathcal{H}}$ ($z =1 $) ~ & $\eta_B$ & ~Br$(\mu\to e\gamma)$~ & ~$m_{\beta\beta}$(eV) ~ & ~NU~~\\
    \hline
      III ~& \,$7\times 10^{-7}$\, &\,$3\times 10^{-3}$\, &  \,-0.93 + 0.26\,$i$\, & \,-1.24  0.22\,$i$\,& \,-0.15 + 0.33\,$i$ \,& ~~2~~ &  5 & $1.23\times 10^{5}$    & ~\,$2.46 \times 10^{-13}$~ & $1.47 \times 10^{-3}$ & 0.004 & ~\checkmark~\\ 
  \hline
  IV~ & \,$7.98\times 10^{-7}$  \,&\, $1.81\times 10^{-3}$\, &\, $-1.73+0.96\,i$ \, &\, -1.84-0.31\,$i$ \, &\,$-1.45+0.75\,i$\, & ~~3~~ & 5 &  57.84  &~ \,$ 6.1\times10^{-10}~$ & $5.98 \times 10^{-13}$ & 0.002& ~\checkmark~\\ 
  \hline
\end{tabular}
\caption{Benchmark choices for the relevant parameters of leptogenesis and the corresponding washout amount which altogether yield the final $\eta_B$, considering a complex $\mathbb{R}$. Unlike the real $\mathbb{R}$ case of table \ref{tab:CIreal}, here CP violation is driven by the phases present in $R$ in addition to the Dirac CP phase. Here also, we have set $M_1=M_2=M_3=1$\,TeV.}
\label{tab:CIcomplex}
\end{center} 
\end{table}
\end{center}
\end{widetext}
Next, we specify the total energy density at any temperature ($T>T_r$) as~\cite{DEramo:2017gpl}
\begin{align}\label{totalrho}
 \rho(T) &= \rho_{rad}(T)+\rho_{\eta}(T)\\
 &=\rho_{rad}(T)\left[1+\frac{g_* (T_r)}{g_* (T)}\left(\frac{g_{*s}(T)}{g_{*s}(T_r)}\right)^{(4+n)/3}\left(\frac{T}{T_r}\right)^n\right]\label{eq:totED}
\end{align}
\noindent From the above equation, it is evident that the energy density of the Universe at any arbitrary temperature ($T>T_r$), is dominated by $\eta$ component. Now, the Friedmann equation, connecting the Hubble parameter with the energy density of the Universe is given by $\mathcal{H}^2 = \frac{\rho}{3M_{\text{Pl}}^2}$ with $M_{\text{Pl}}= 2.4 \times 10^{18}$ GeV being the reduced Planck mass. At temperature higher than $T_r$ with the condition $g_*(T) = \bar g_*$ which can be considered to be some constant, the Hubble rate can approximately be recasted into the following form~
\begin{align}\label{eq:mod-hubl}
 \mathcal{H}(T) &\approx \frac{\pi\bar g_*^{1/2}}{3\sqrt{10}} \frac{T^2}{M_{\text{Pl}}}\left(\frac{T}{T_r}\right)^{n/2}, ~~~~ ({\rm with ~~}T \gg T_r),\\ \nonumber
 &=\mathcal{H}_R(T)\left(\frac{T}{T_r}\right)^{n/2}, 
\end{align}
\noindent where $\mathcal{H}_R(T)\simeq 0.33~\bar{g}_*^{1/2}\frac{T^2}{M_{\rm Pl}}$, the standard Hubble rate for a radiation dominated Universe. In case of SM, $\bar g_*$ can be identified with the total SM energy degrees of freedom $g_*\text{(SM)} \simeq 106.75$ at temperatures above 100 GeV. In the present analysis we anticipate that if non-trivial domination of $\eta$ species persists till late time at early Universe, it might be possible to delay the inverse decays and subsequently reduce the amount of washout. {\color{black}It is worth to mention here, that 
for a non-standard cosmology, the analytical expression of $\eta_B$ in Eq.(\ref{Eq:bau}) does not hold and one needs to solve the full set of BEQs as given in Eq.(\ref{beq1}) and Eq.(\ref{beq1}) in order to find a correct prediction of the final amount of lepton asymmetry. Importantly, the Hubble parameter $\mathcal{H}(T)$, present in Eq.(\ref{beq1}) and Eq.(\ref{beq1}) takes the form of Eq.(\ref{eq:mod-hubl}) while we solve the BEQs considering non-standard cosmology.}

\begin{center}
\begin{figure*}[htb!]
    \includegraphics[height=4.2cm,width=4.2cm]{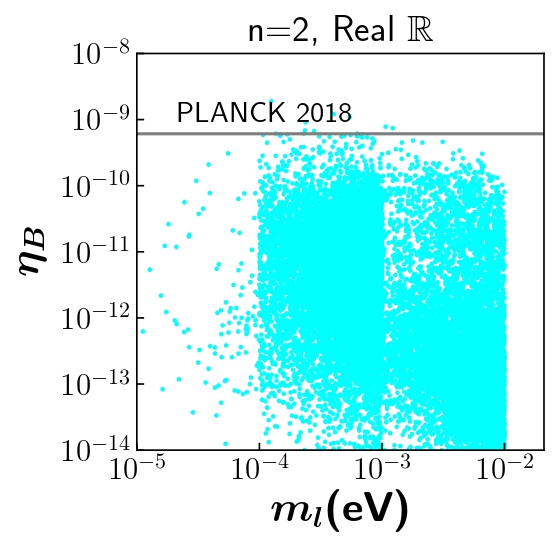}~
    \includegraphics[height=4.2cm,width=4.2cm]{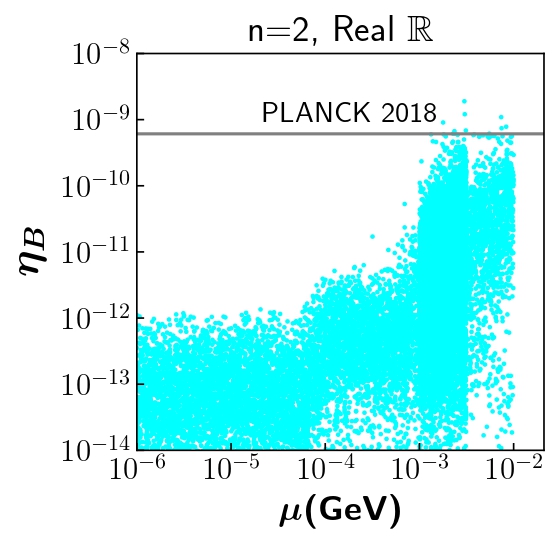}
        ~\includegraphics[height=4.2cm,width=4.2cm]{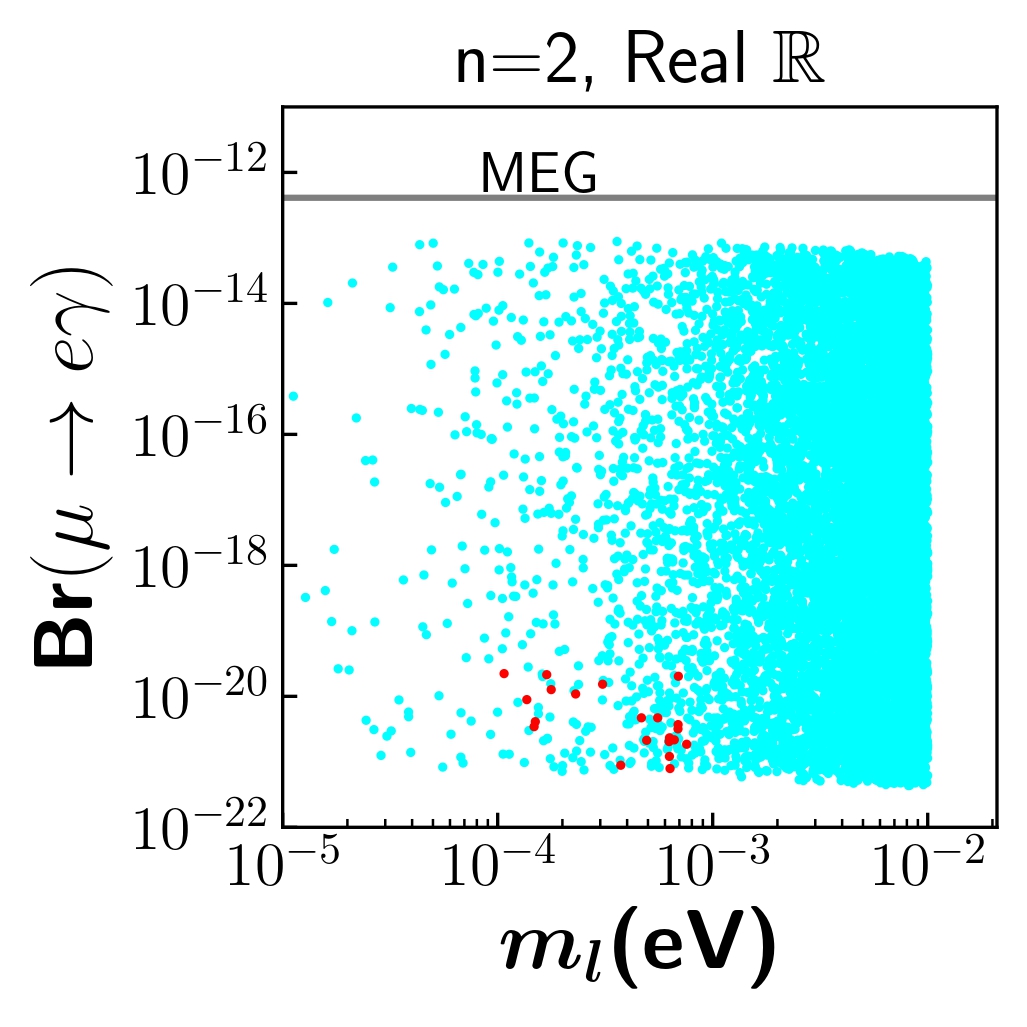}~            \includegraphics[height=4.2cm,width=4.2cm]{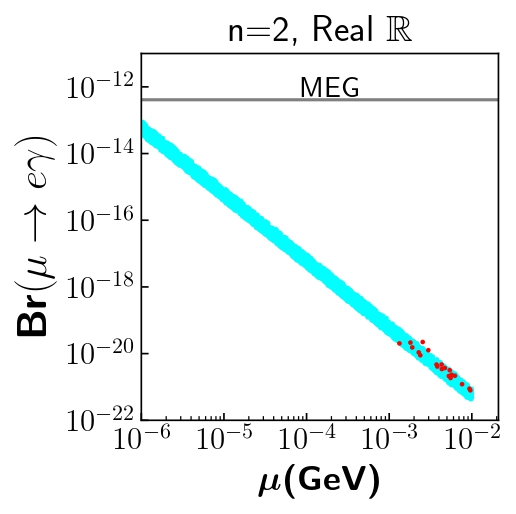}
\caption{In the left estimate of $\eta_B$ is shown as a function of the lightest neutrino mass ($m_l$) (first left) and $\mu$ (second left) considering $\mathbb{R}$ real. Right column shows the prediction for Br$(\mu\to e\gamma)$ w.r.t. the same parameters as in the left. Here we have set $(n,T_r)=(2,5\,{\rm MeV})$. The grey horizontal line in left and right columns indicate the experimentally observed value of $\eta_B$ and the present limit on Br$(\mu\to e\gamma)$ by MEG \cite{MEGII:2018kmf} respectively. The red dots are viable points that give rise to correct amount of baryon asymmetry, without violation of  Br$(\mu\to e\gamma)$ and non-Unitarity bounds.}
\label{fig:RSreal1}
 \end{figure*}
\end{center}
\begin{center}
\begin{figure*}
    \includegraphics[height=4.52cm,width=4.52cm]{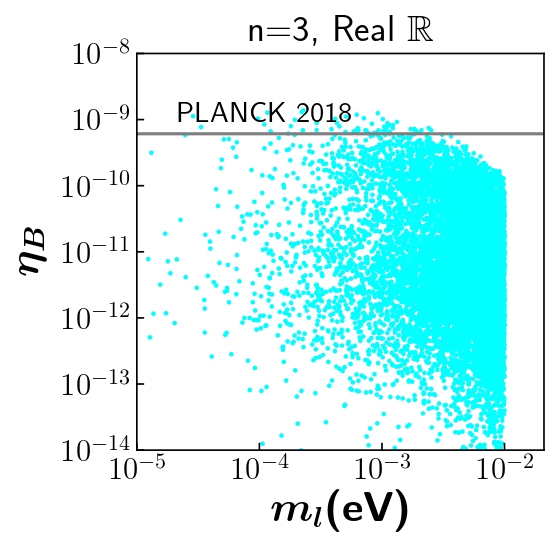}
    \includegraphics[height=4.52cm,width=4.52cm]{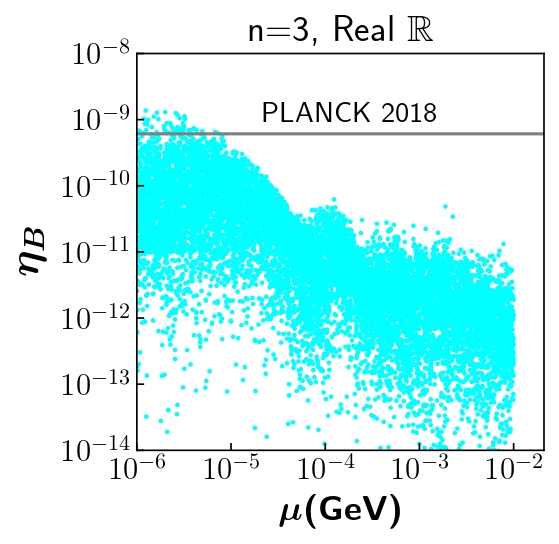}
        \includegraphics[height=4.52cm,width=4.52cm]{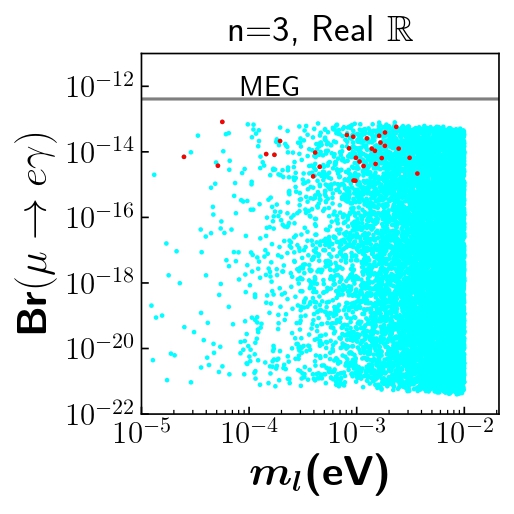}
         \includegraphics[height=4.52cm,width=4.52cm]{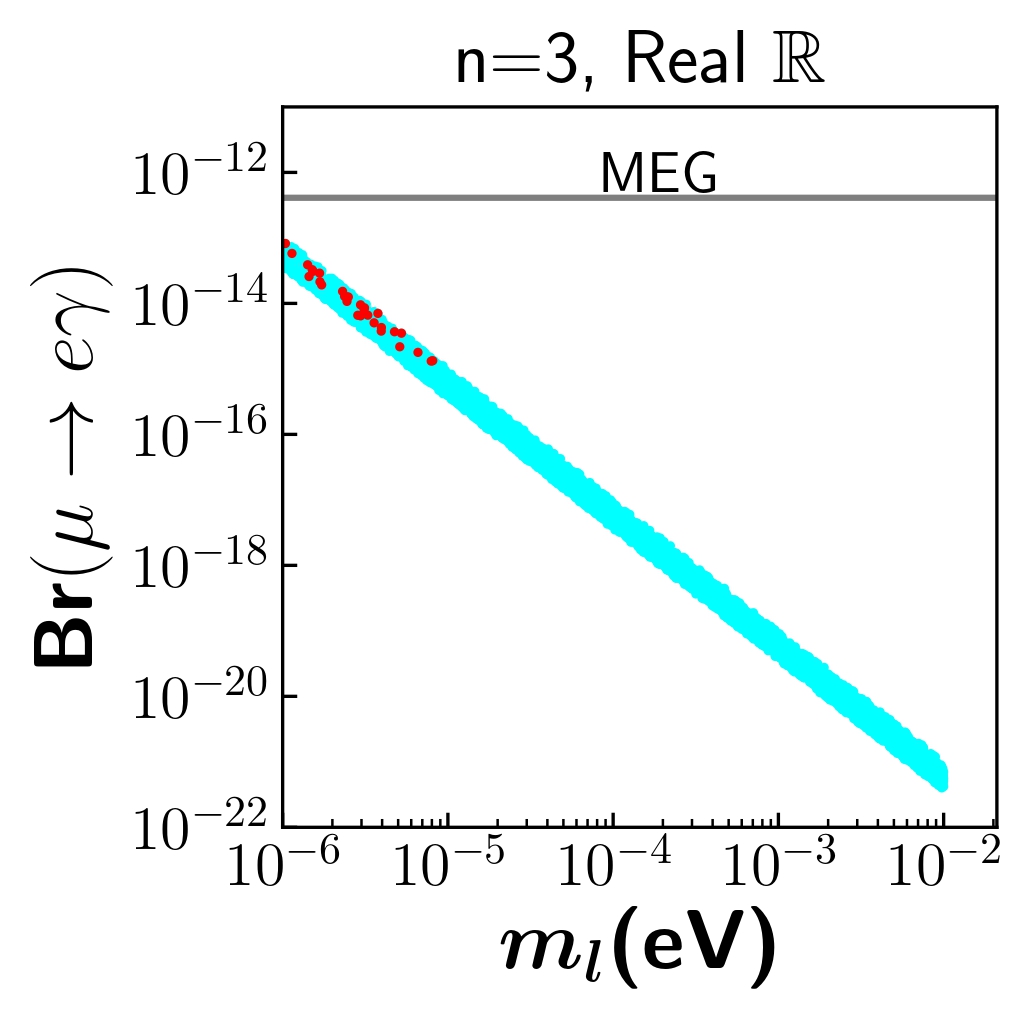}
\caption{Same as in Fig.\,\ref{fig:RSreal1}. Here we have set $(n,T_r)=(3,5\,{\rm MeV})$. The grey horizontal lines in the left and right columns indicate the observed value of $\eta_B$ \cite{ade2016planck} and the present sensitivity on Br$(\mu \to e\gamma)$ set by MEG \cite{Aubert:2009ag,Adam:2013mnn,MEG:2016leq} respectively. The red dots are viable points that give rise to correct amount of baryon asymmetry, without violation Br$(\mu\to e\gamma)$ and non-unitarity bounds.}
\label{fig:RSreal2}
\end{figure*}
\end{center}

First, in Table\,\ref{tab:CIreal}, we tabulate two benchmark points (for real $\mathbb{R}$) that successfully yield the observed baryon asymmetry of the Universe, satisfying the neutrino oscillation data at the same time.  For BP-CA I, we have assumed the presence of kination domination ($n=2$) prior to BBN era while in BP-CA II, presence of a fluid ($n=3$) in early Universe is considered which redshifts faster than kinaton. In first two columns of Figs.\,\ref{fig:C1BPIreal} and \ref{fig:C1BPIIreal}, we have shown the evolution of $\eta_{B-L}$ as a function of temperature for both the benchmark points considering standard and non-standard cosmology respectively. In standard case ($n\simeq 0$), the produced lepton-asymmetry gets substantially suppressed due to strong wash out effects, predominantly from inverse decay. The impact of huge wash-out in reducing the produced lepton asymmetry is evident from bottom last two columns of Figs.\,\ref{fig:C1BPIreal}and \ref{fig:C1BPIIreal} where the ratio of thermally averaged decay (inverse decay) rate and Hubble parameter is plotted as a function of temperature. Interestingly, the strength of wash-out gets heavily diluted (see bottom right panel of Figs.\,\ref{fig:C1BPIreal} and \ref{fig:C1BPIIreal}) when we consider the presence of non-standard cosmology in form of kinaton or faster than kinaton dominated Universe. Such suppression in the amount of washout owing to the presence of non-standard cosmology assists in obtaining the right order of baryon asymmetry abundance. Considering a complex $\mathbb{R}$, we also provide two suitable benchmark points (BP-CA III and BP-CA IV, see table\,\ref{tab:CIcomplex}) that obey the neutrino oscillation data. Similar to the case with a real $\mathbb{R}$, here also we have considered the presence of $\eta$ fluid at early Universe with different set of $(n,T_r)$ values. Despite BP-CA III being consistent with the bounds from NU and Br$(\mu\to e\gamma)$ constraints, it falls short in providing the desired amount of $\eta_B$ due to insufficient suppression of washout strength. In fact, we have observed that the $n=2$ case for a complex $\mathbb{R}$ does not work in rescuing the ISS parameter space for leptogenesis regardless of the value of $T_r$ and thus we refrain from discussing this particular case further.

Next, we proceed to perform a random scan over the relevant parameters that are involved in controlling the value of $\eta_B$. For the purpose, we fix the following ranges,
\begin{align}
     5\times 10^{-7}{\rm\, GeV}& \leq \mu\leq 10^{-2}{\rm\,GeV},\nonumber\\
    10^{-5}{\rm~eV}&\leq m_l \leq 10^{-2}{\rm eV},\nonumber\\
     -4\pi&\leq |x|,|y|,|z|\leq 4\pi\,.
\end{align}
and set $M_1=M_2=M_3=1$\,TeV.

We first consider $\mathbb{R}$ as real and attempt to find out the parameter space that can generate adequate amount of baryon asymmetry in the Universe. In the first two columns of Fig.\ref{fig:RSreal1} and Fig.\ref{fig:RSreal2}, we have shown the obtained values for $\eta_B$ as function of $m_l$ and $\mu$ respectively. In the last two columns of Fig.\ref{fig:RSreal1} and Fig.\ref{fig:RSreal2} the predictions for Br$(\mu\to e\gamma)$ are presented as with respect to $m_l$ and $\mu$ respectively along with current sensitivity of MEG\,\cite{Aubert:2009ag,Adam:2013mnn,MEG:2016leq}. As earlier stated, a faster expansion of the Universe, characterised by suitable choices of $(n,T_r)$ assists in suppressing the strength of washout process. 
This feature facilitates in rescuing part of the ISS parameter space where the PLANCK 2018 bound on the baryon asymmetry can be satisfied which is otherwise not possible in a conventional radiation dominated Universe. In Fig.\,\ref{fig:RSreal1}, the non-standard cosmological parameters are set as $(n,T_r)=(2,5\,{\rm MeV})$. Indeed there exists a few viable points (marked by red dots in last two column of Fig.\ref{fig:RSreal1}), that yield the expected amount of baryon asymmetry with $\mu\sim\mathcal{O}(10^{-3})-\mathcal{O}(10^{-2})$ keV. Note that for a real nature of $\mathbb{R}$, $\delta_{\rm CP}$ being the only source of CP violation, the non-Unitarity constrains does not restrict the parameter space. Also, the prediction for Br$(\mu\to e\gamma)$
comes out to be much smaller than the present sensitivity of MEG experiment. We repeat the same analysis in Fig.\,\ref{fig:RSreal2} for a non-standard Universe ($n=3,T_r=5\,{\rm MeV}$) dominated by a fluid that redshifts faster than kinaton. We notice that such choice of ($n,T_r$) results into a somewhat enhanced parameter space preferring relatively lower $\mu\sim \mathcal{O}(1)-\mathcal{O}(10)$ keV. Due to the preference over smaller $\mu$, here the estimate of Br$(\mu\to e\gamma)$ turns out to be relatively closer to the MEG experimental sensitivity compared to the earlier case.

\begin{centering}
\begin{figure*}[htb!]
    \includegraphics[height=4.4cm,width=4.52cm]{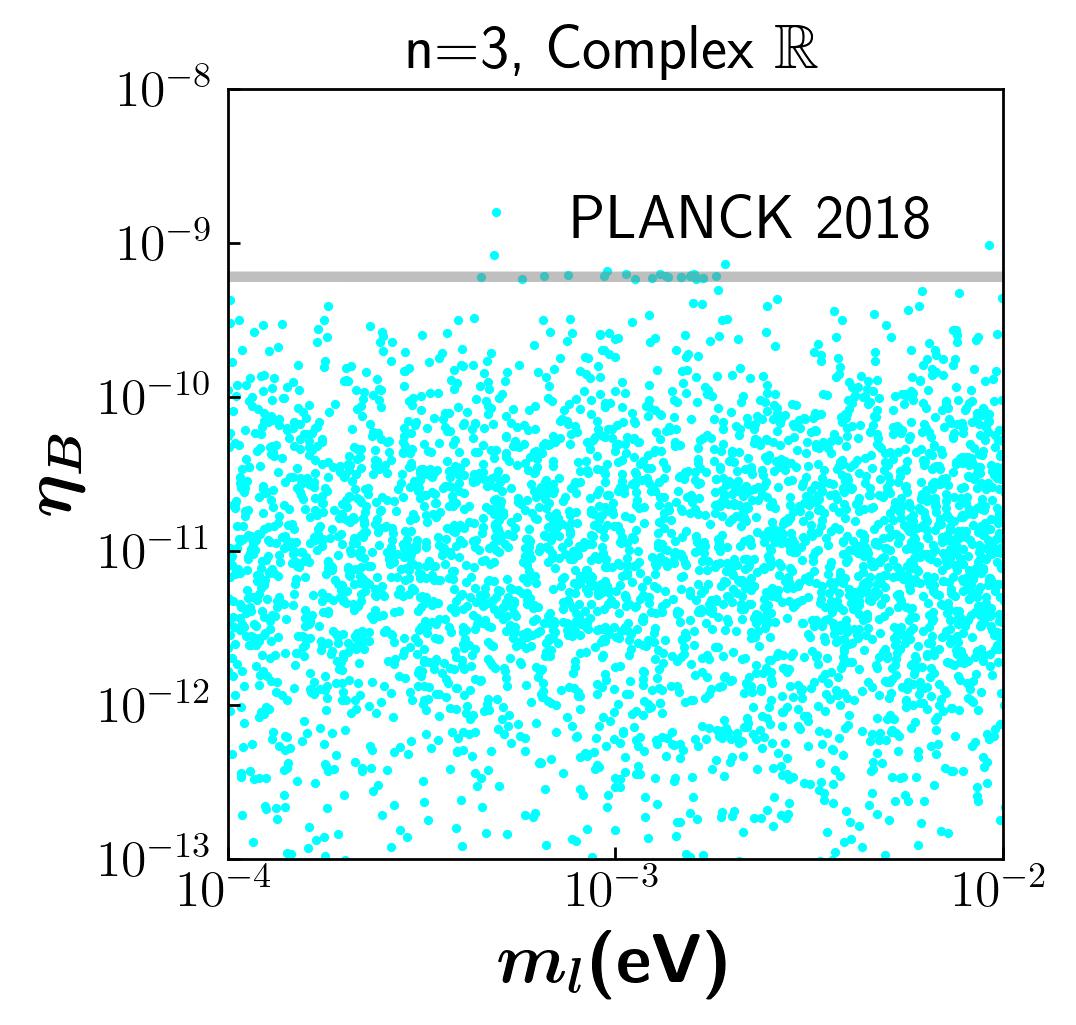}
    \includegraphics[height=4.4cm,width=4.52cm]{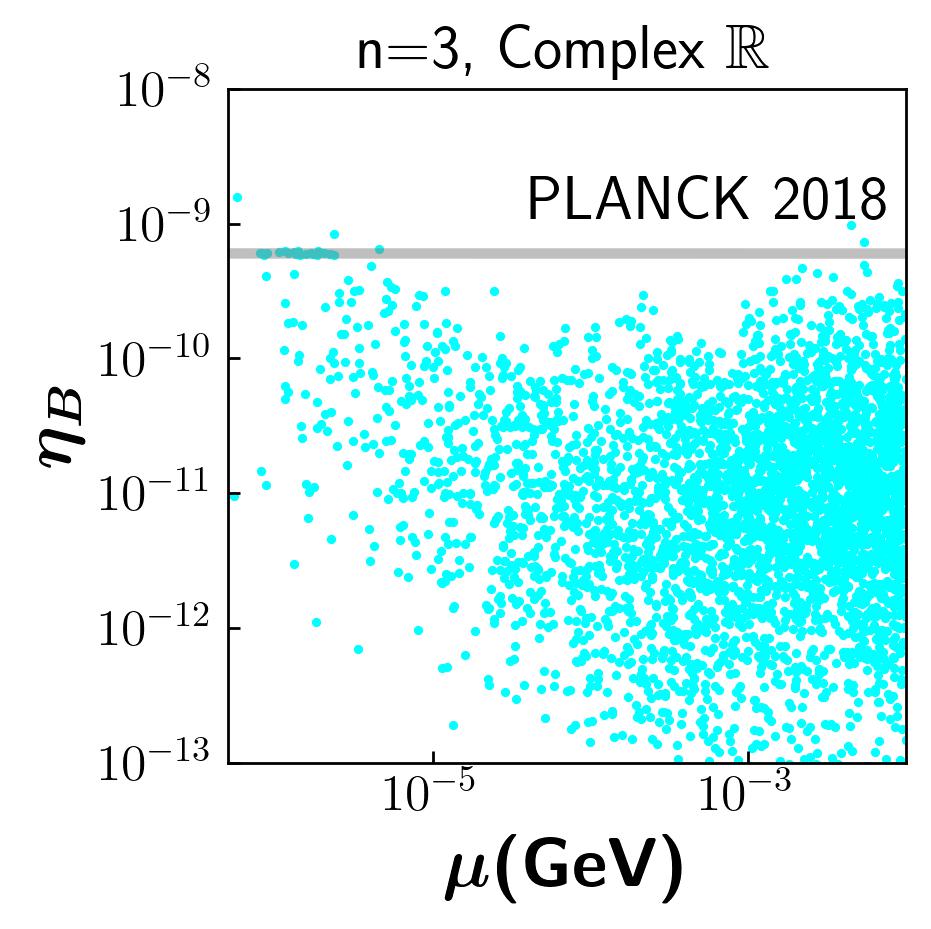}
     \includegraphics[height=4.4cm,width=4.52cm]{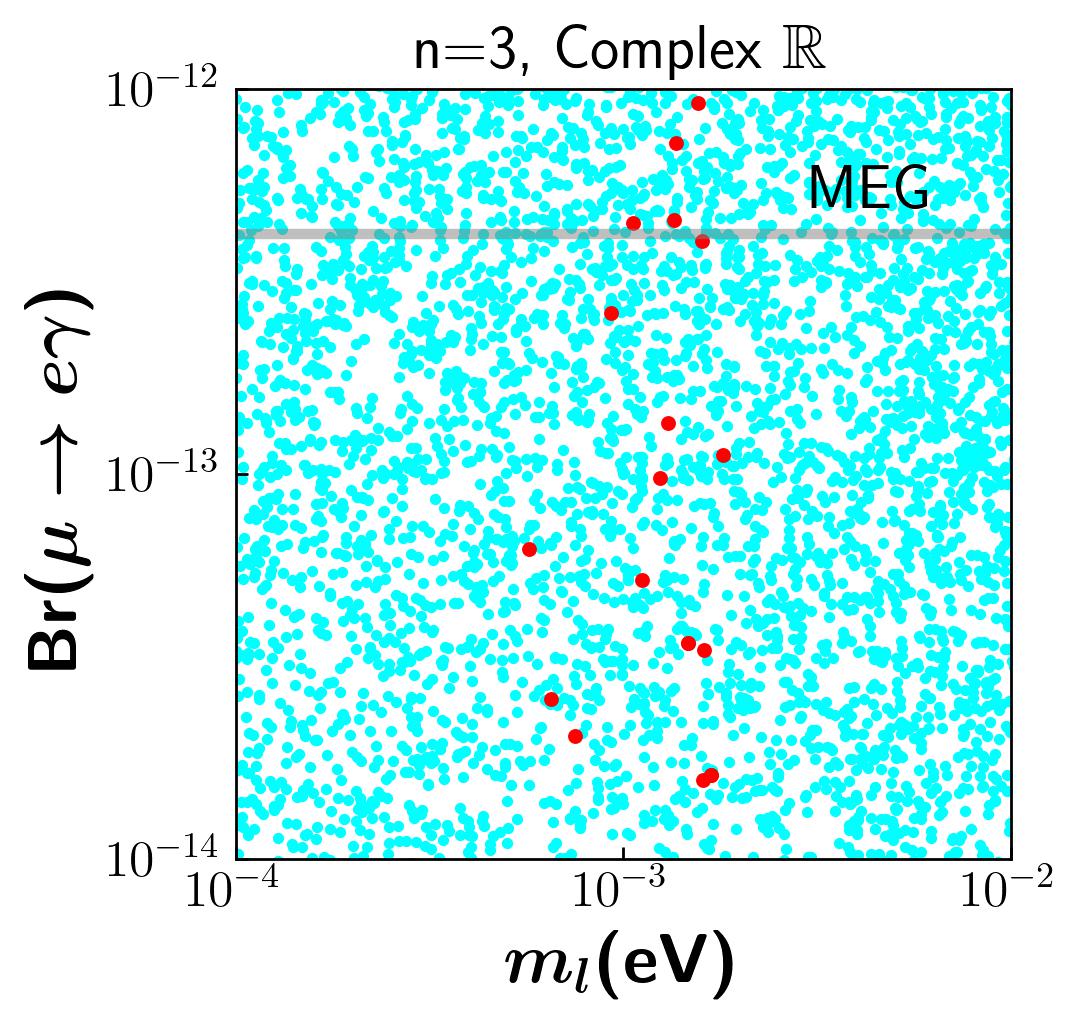}
    \includegraphics[height=4.4cm,width=4.52cm]{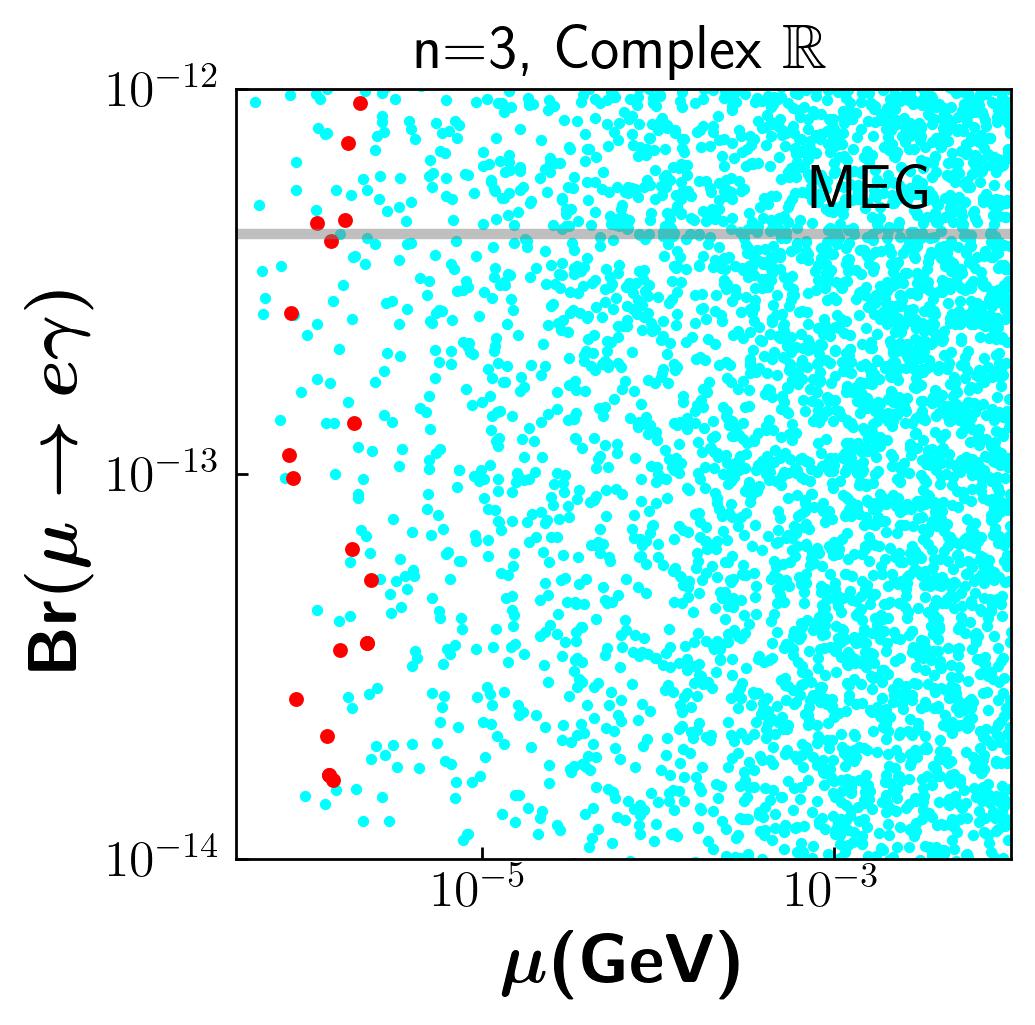}
\caption{Viable parameter space for complex $\mathbb{R}$ when $n=3$. In first two columns, we show the estimate of $\eta_B$ as function of $\mu$ and $m_l$ respectively. Last two columns display the estimate of Br$(\mu\to e\gamma$) w.r.t. $\mu$ and $m_l$ respectively. The cyan colored points show the yield for baryon asymmetry and the branching ratio after imposition of NU constraints.
The red dots in the right columns show the relevant ranges of $\mu$ and $m_l$ which satisfy the baryon asymmetry criteria.}
 \label{fig:RScomplex3}
 \end{figure*}
\end{centering}
\begin{figure*}[htb!]
\begin{center}
        \includegraphics[height=4.5cm,width=4.5cm]{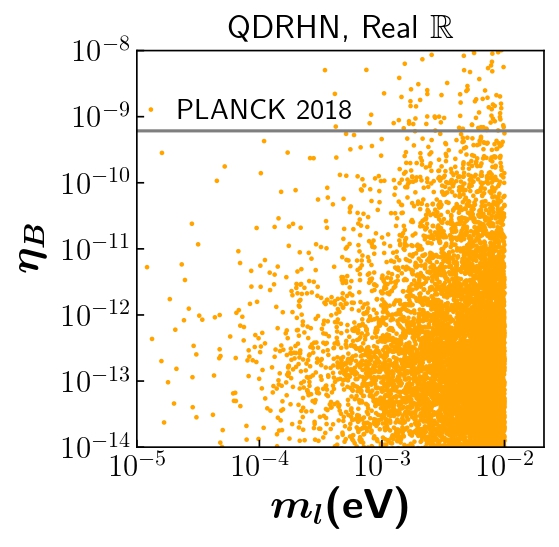}
        \includegraphics[height=4.5cm,width=4.5cm]{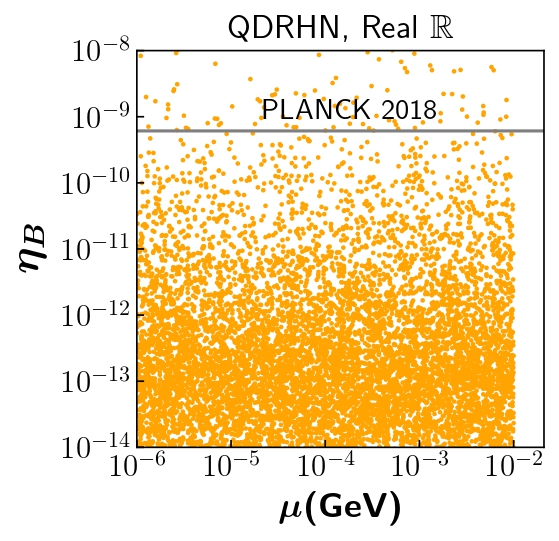}
        \includegraphics[height=4.5cm,width=4.5cm]{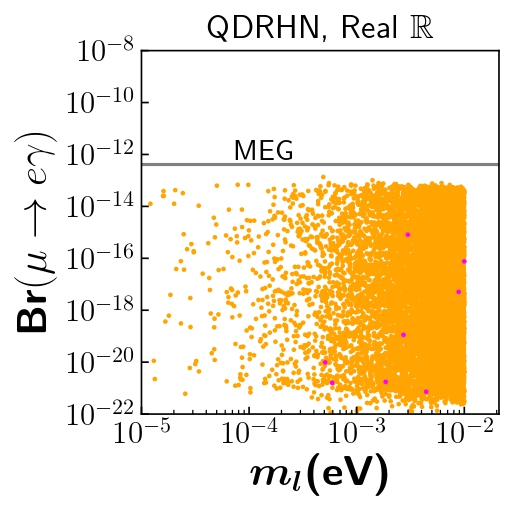}   
  \includegraphics[height=4.5cm,width=4.5cm]{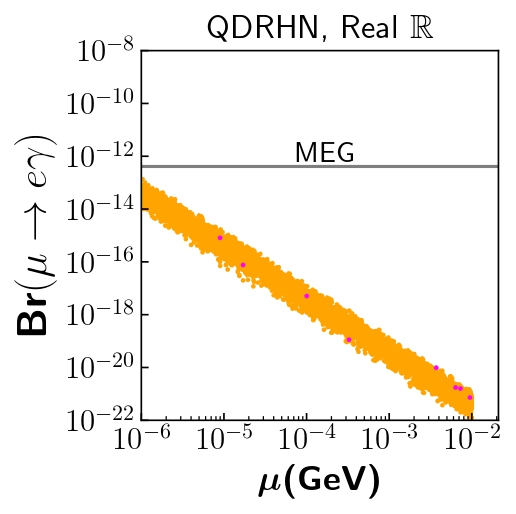} 
\caption{The left two columns show baryon to photon ratio for real $\mathbb{R}$ as a function of the LNV scale ($m_l$) and the lightest neutrino mass
($\mu$) when the heavy RHN states are non-degenerate, keeping $M_1 = 1$TeV, $M_2= 1.01$ TeV, $M_3= 3$ TeV. Here we have varied $\delta_{\rm CP}$ from $\frac{\pi}{6}$ to $\frac{3\pi}{2}$. The right columns correspond to the respective branching ratio w.r.t. $\mu$ and $m_l$. For the real $\mathbb{R}$ the ISS parameter space
offering leptogenesis is not restricted by non-Unitarity of lepton mixing. The gray band in the left plots reports the PLANCK bound on $\eta_B$, whereas the one in the RHN plots indicate the present sensitivity on the Br$(\mu\to e\gamma)$ reported by MEG. The magenta points in the right columns represent those which satisfy PLANCK bound on $\eta_B$.}
\label{fig:real_NDRHN}
\end{center}
 \end{figure*}

We have clarified earlier that for a real $\mathbb{R}$, the Dirac CP phase ($\delta_{\rm CP}$) is the only CP violating parameter that leads to ample amount of remnant baryon asymmetry in the Universe. When $\mathbb{R}$ is complex, one has additional CP violating sources. We analyse the case for $n=3$ for a complex $\mathbb{R}$ in Fig.\,\ref{fig:RScomplex3}. We notice that such choice of ($n,T_r$) results into a few allowed points (marked by red dots in last two columns of Fig.\,\ref{fig:RScomplex3}), favored for explaining the observed value of $\eta_B$, surviving the strong non-Unitarity and Br$(\mu\to e \gamma)$ bounds. We also observe that the satisfaction of PLANCK 2018 bound prefers $\mathcal{O}(1)\,{\rm keV}\lesssim\mu\lesssim \mathcal{O}(10)\,{\rm keV}$. Due to the preference over smaller $\mu$, the estimate of Br$(\mu\to e\gamma)$ for baryon asymmetry satisfied points turns out to be closer to the MEG experimental sensitivity.

\subsection{Case B: Increasing lepton asymmetry}
\begin{figure*}[htb!]
\begin{center}
        \includegraphics[height=4.5cm,width=4.5cm]{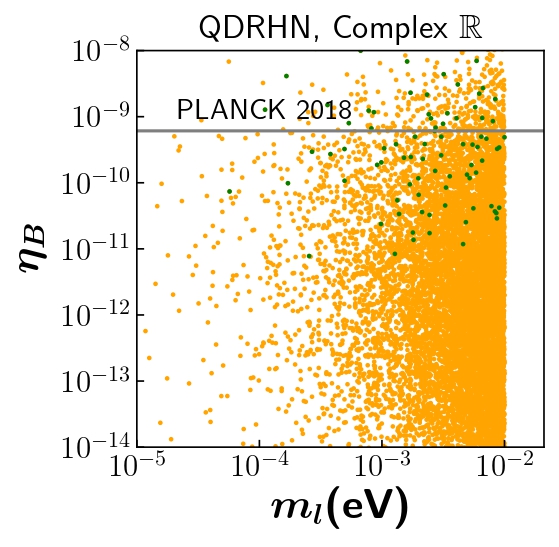}
        \includegraphics[height=4.5cm,width=4.5cm]{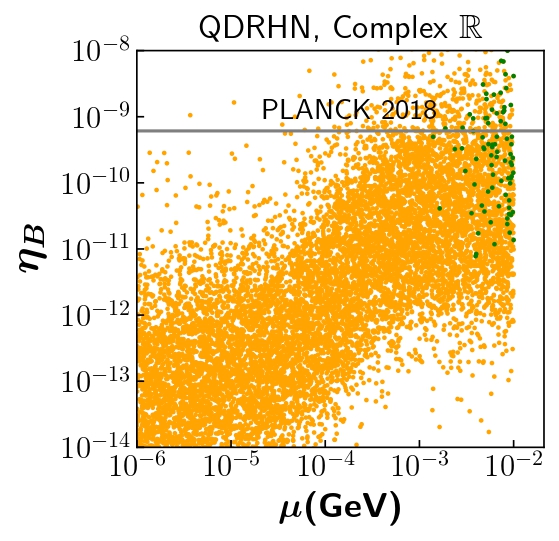}
        \includegraphics[height=4.5cm,width=4.5cm]{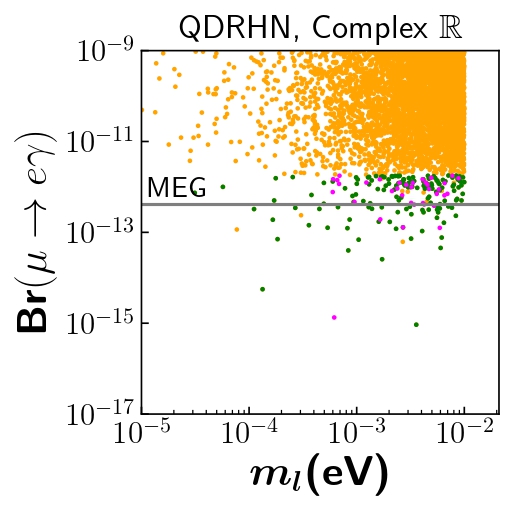}   
  \includegraphics[height=4.5cm,width=4.5cm]{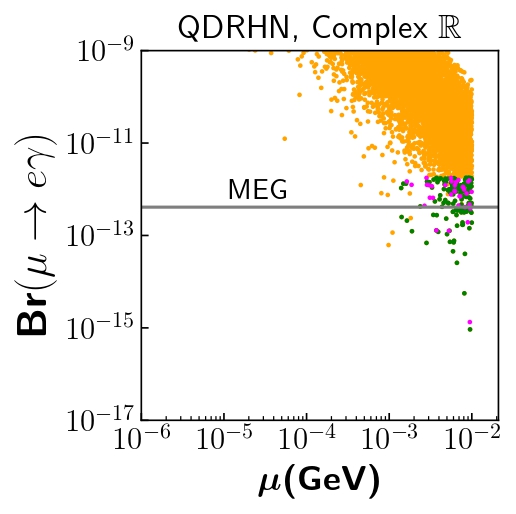} 
\caption{Same as in Fig.\,\ref{fig:real_NDRHN} with exception complex $\mathbb{R}$. For complex $\mathbb{R}$, a part of the ISS parameter space is ruled out by simultaneous imposition of baryon asymmetry and NU criteria. The viable parameter space satisfying these two constraints is shown by green points. The magenta colored points in the right columns indicate those which satisfy both $\eta_B$ and NU constraints simultaneously.}
\label{fig:complex_NDRHN}
\end{center}
 \end{figure*}

\begin{figure*}
\centering
 \includegraphics[height=4.5cm,width=4.25cm]{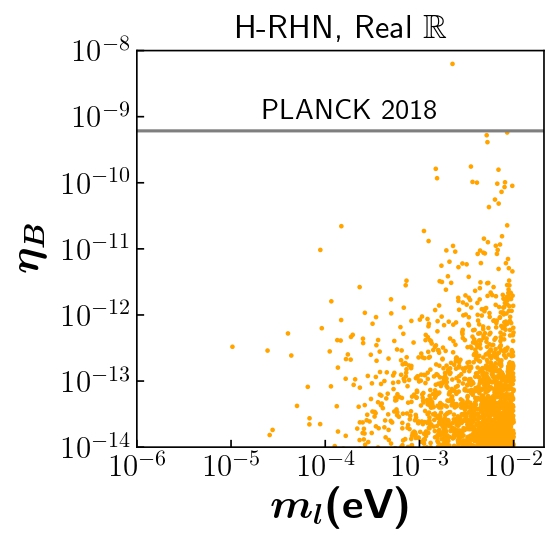}
       \includegraphics[height=4.5cm,width=4.25cm]{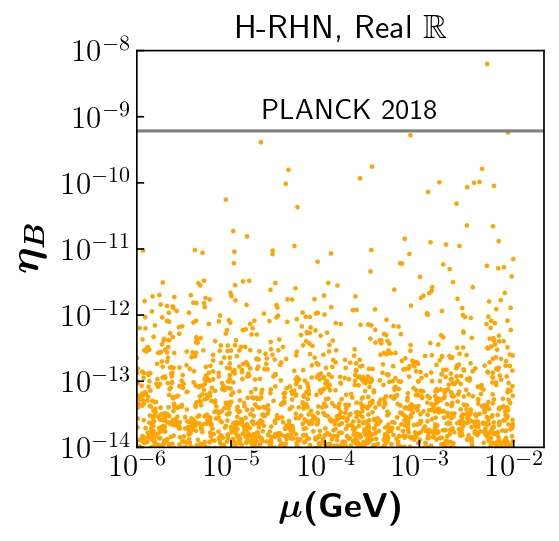}
\includegraphics[height=4.5cm,width=4.25cm]{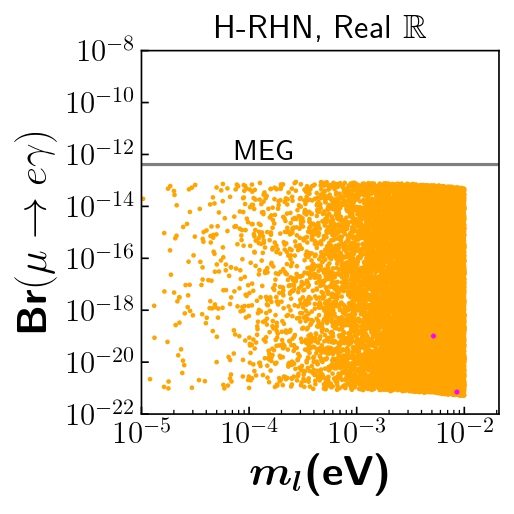}
\includegraphics[height=4.5cm,width=4.25cm]{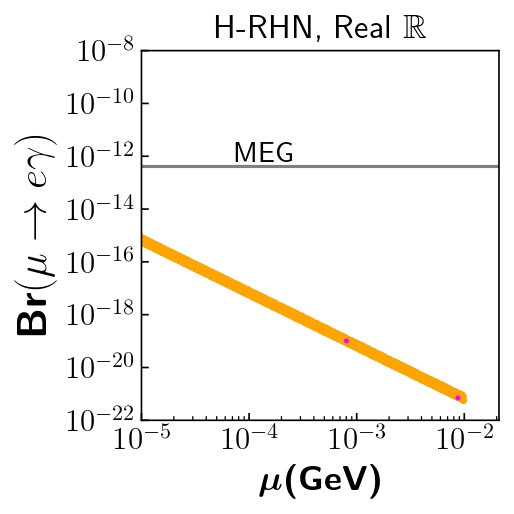} 
\caption{Same as in Fig.\,\ref{fig:real_NDRHN} with hierarchical RHN masses, e.g. $M_1 = 1$\,TeV, $M_2= 2$\,TeV, $M_3= 3$\,TeV. The magenta colored points in the right columns indicate those which satisfy both $\eta_B$ and NU constraints simultaneously.}
\label{fig:real_NNRHN}
\end{figure*}

\begin{figure*}
\begin{center}
\includegraphics[height=4.5cm,width=4.25cm]{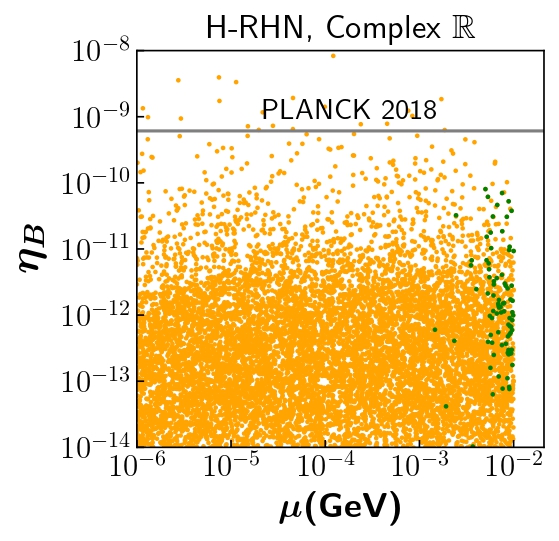}
        \includegraphics[height=4.5cm,width=4.25cm]{etab_ml_complexhrhn.jpg}    
\includegraphics[height=4.5cm,width=4.25cm]{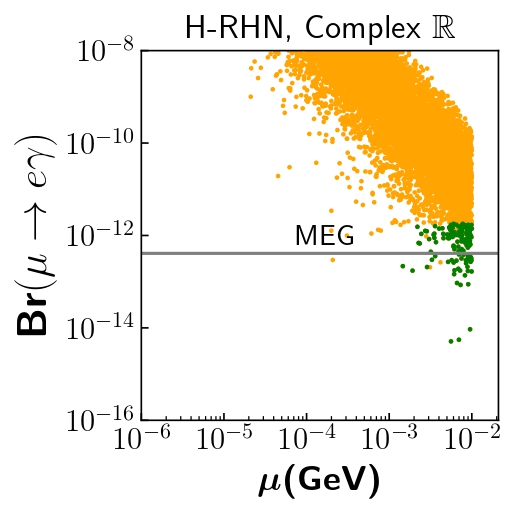}
\includegraphics[height=4.5cm,width=4.25cm]{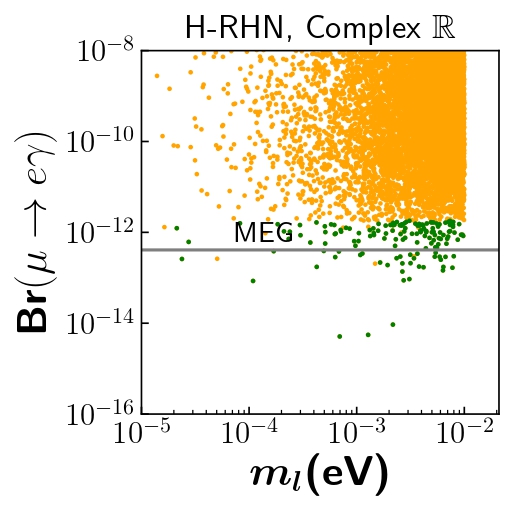}
\caption{Same as the caption of Fig.\,\ref{fig:complex_NDRHN} with hierarchical RHN masses, e.g. $M_1=1$\,TeV, $M_2=2$\,TeV, $M_3=3$\,TeV. Here, we do not get any points which simultaneously satisfy the constraints from baryon asymmetry and NU. The green points which respect the NU bounds do not yield the observed baryon to photon ratio.}
\label{fig:complex_NNRHN}
\end{center}
\end{figure*}

As mentioned in Sec.\ref{sec:lep}, the ISS parameter space for the survival of final lepton asymmetry can be retrieved by lifting the degeneracy among the elements in the $M_R$ matrix. In this section we present such investigation subject to both real and complex choices of $\mathbb{R}$. For the computation of lepton asymmetry we have considered two set of choices, $M_1 = 1$\,TeV, $M_2= 1.01$\,TeV, $M_3= 3$\,TeV and $M_1 = 1$\,TeV, $M_2= 2$\,TeV, $M_3= 3$\,TeV respectively. Like the previous cases, we have varied here the LNV scale of ISS model as $\mu = 10^{-6} \,-\, 10^{-2}$\,GeV and the lightest neutrino mass ($m_l$) from $10^{-5}\,-\,10^{-2}$\,eV. We have shown the corresponding results in Figs.\,\ref{fig:real_NDRHN}, \ref{fig:complex_NDRHN}, \ref{fig:real_NNRHN}, and \ref{fig:complex_NNRHN} respectively associated with the real and complex $\mathbb{R}$ cases and the two choices made for $(M_1,M_2,M_3)$ as earlier specified. Inclusion of non-degeneracy in $M_R$,  corresponding to different generations of heavy sterile states helps to avoid the partial cancellation between the lepton asymmetries associated with the pseudo-Dirac states forming a pair. This in turn leads to a sufficient yield of total lepton asymmetry contributed by all the pseudo-Dirac states. The reason behind the aforementioned choice of two sets of ($M_{1,2,3}$) is to see the influence of the appropriate amount of degeneracy in saving the ISS parameter space. For a better clarity on this issue, in table \ref{tab:caseB} and \ref{tab:Yukawa} we mention the benchmark points and the corresponding outcomes, relevant to the case B. The respective Yukawa coupling matrices obtained for each individual benchmark points mentioned in table \ref{tab:caseB} have been presented in table \ref{tab:Yukawa}. The corresponding mass inputs for the pseudo-Dirac states are $M_1 = 1$\,TeV, $M_2= 1.01$\,TeV, $M_3= 3$\,TeV. Considering a real $\mathbb{R}$, in the left columns of Fig.\,\ref{fig:real_NDRHN} we show baryon to photon ratio as a function of $\mu$ and $m_l$ for the first set of RHN mass states mentioned above. In the right columns we present the branching ratio for $\mu \to e\gamma$ w.r.t. $\mu$ and $m_l$.

Fig.\ref{fig:complex_NDRHN} represents the baryon to photon ratio (left) and branching ratio for $\mu \to e\gamma$ (right) same for the complex $\mathbb{R}$ with the first choice of RHN mass states. For this case we notice that the some of the green points which satisfy NU constraints along with the desired baryon asymmetry, also yield the Br($\mu \to e\gamma$) which is close to the present and future sensitivity. However, the final points allow only larger value of $\mu$ and the lightest neutrino mass. On the 
other hand we show these results for the second choice of RHN masses in Fig.\,\ref{fig:real_NNRHN} and Fig.\,\ref{fig:complex_NNRHN} respectively for the real and complex $\mathbb{R}$. From these four figures it is clearly evident that an adequate degeneracy among the RHN mass states are crucial to retrieve the ISS parameter space for successful leptogenesis. From Fig.\,\ref{fig:real_NNRHN} and Fig.\,\ref{fig:complex_NNRHN} we learn that, a larger hierarchy among the RHN mass states can change the leptogenesis prediction of ISS model significantly. Similar conclusion can also be drawn from Fig.\,\ref{fig:real_NDRHN} and Fig.\,\ref{fig:complex_NNRHN}. For a hierarchical spectrum of RHNs the number of points crossing the PLANCK bound on baryon to photon ratio is very less, as it is evident in the respective figures. The reason being, for strictly hierarchical RHN mass states it is difficult to generate adequate amount of lepton asymmetry which can finally compete against the huge washout (with $K \sim \mathcal{O}(10^{7})$). While investigating this case we imposed relevant constraints associated with the non-Unitarity of the lepton mixing. This is evinced especially in Fig.~\ref{fig:complex_NDRHN} and Fig.\,\ref{fig:complex_NNRHN}, by the orange points, which are excluded by the bounds on non-Unitary mixing. The NU mixing matrix elements are controlled by several electro-weak interaction observables mentioned in \cite{Fernandez-Martinez:2016lgt}. The exclusion of ISS parameter space, in terms of $\mu$ has taken place even through the NU constraint, in addition to the baryon asymmetry criteria. It is because of the complex $\mathbb{R}$ structure which has hyperbolic dependency on the mixing angles, thus making the light-heavy mixing very large. Which, however is not the case for real $\mathbb{R}$. For the real case, thus NU does not restrict the ISS parameter space, whether complex $\mathbb{R}$ does as evident in Fig.~\ref{fig:complex_NDRHN} and Fig.~\ref{fig:complex_NNRHN}. Therefore the green points are the final representative parameter space of ISS offering successful leptogenesis which can be indirectly probed through LFV search.    

\begin{widetext}
\begin{center}
\begin{table*}[h]
\addtolength{\tabcolsep}{-1.4pt}
  \renewcommand*{\arraystretch}{1.4}
\begin{tabular}{| c | c | c | c | c | c | c | c | c | c | c | c | c | c |}
  \hline
\,BP-CB\, &$\mu$\,(GeV)& $m_l$\,(eV) &~~ $x$~~ &~~ $y$~~ & ~~$z$~~  & $\sum_{i=1}^6\epsilon_i^e$ & $\sum_{i=1}^6\epsilon_i^\mu$\, & $\sum_{i=1}^6\epsilon_i^\tau$ & $\eta_B$ & Br($\mu \to e\gamma$) & $m_{\beta \beta}$(eV)\, & \,NU \,\\
    \hline
      I& \,$0.0098$\, &\,$0.006$\, &  \,$3.92 + 4.34 i$\, & \,$1.9 + 2i$\, & \,$2.85 + 0.1 i$\, \,& \,$-0.001$\, & \,$0.004$\, & \,$-0.003$\,    & \,$5.9 \times 10^{-10}$\, &\, $8.59 \times 10^{-13}$\, & $ 0.006 $ & ~\checkmark~ \\  
       \hline
      II& \,$0.009$\, &\,$0.006$\, & \, $3.49 + 4.2 i$\, &\, $3.5 + 2 i$\, &\, $2.7 + 0.1 i$ \,& \,$-6.4\times 10^{-7}$\, & \,$1.4\times10^{-5}$\, & \,$-1.3 \times 10^{-5}$\, & \,$2.12 \times 10^{-10}$\, &\, $3.97 \times 10^{-13}$\, & \,0.006\, &  ~\checkmark~  \\ 
      \hline
      III & \,$6.4\times 10^{-5}$\, & \,0.0059\, & 3.55 & 2.44 & 3.82\,& $2.6\times 10^{-5} $& $-2.19 \times 10^{-5}$ & $-4.7\times 10^{-6}$ & $ 5.27 \times 10^{-10}$ & $1.3 \times 10^{-17}$ & 0.006 & ~\checkmark~  \\ 
      \hline
      IV& \,$2.2\times10^{-6}$\,& \,$0.0078$\, &\,3.2\, & 2.89  & 2.66 \,& $-8.2\times10^{-6}$ & $3.5 \times10^{-6}$ &  $4.6 \times10^{-6}$  & \,$5.7 \times 10^{-10}$ & $1.17 \times 10^{-14}$ & 0.007& ~\checkmark~ \\ 
  \hline
\end{tabular}
\caption{Benchmark points (I) and (II) representing the parameter space subject to Fig.\,\ref{fig:complex_NDRHN} and Benchmark points (III) and (IV) Fig.\,\ref{fig:real_NDRHN}.}
\label{tab:caseB}
\end{table*}
\end{center}
\end{widetext}

\begin{figure*}
\includegraphics[height=4.5cm,width=6.2cm]{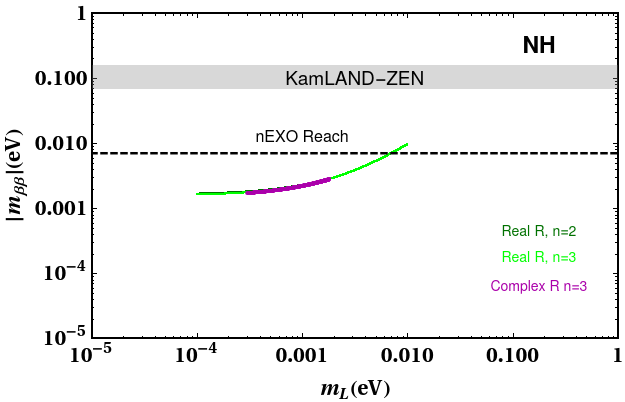}~~~~~~~~~~~~~\includegraphics[height=4.5cm,width=6.2cm]{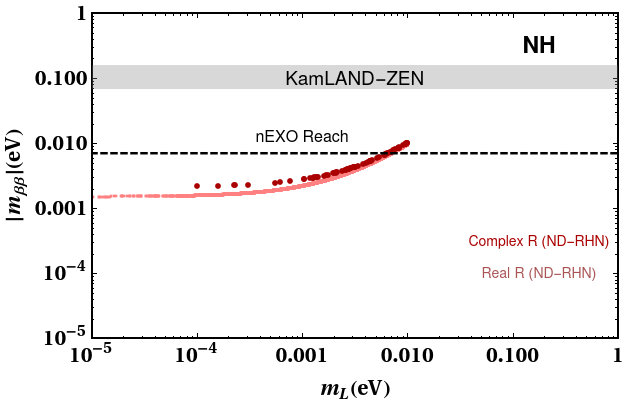}
\caption{Shows the effective neutrino mass predicted w.r.t. the constrained lightest neutrino mass. The figures provide explanations of the implications associated with various color codes, enabling a clear differentiation between each case. The lightest neutrino mass gets restriction from the criteria $\eta_B = (6 - 6.18)\times 10^{-10}$. The grey band indicates the limit on $m_{\beta\beta}$ set by the KamLAND-ZEN experiment \cite{KamLAND-Zen:2016pfg}, which falls in the range $61-165$ MeV. The black dashed horizontal line indicates the future sensitivity set by nEXO \cite{nEXO:2018ylp} on $m_{\beta \beta}\,=\,0.007$\,eV. }
\label{fig:ndbd}
\end{figure*}

\begin{table}[]
\small
    \centering
    \begin{tabular}{|c|c|}
    \hline
    BP-CB & $Y_\nu^{6\times 3}$\\
    \hline
        I & $10^{-3}\left(
\begin{array}{ccc}
 0.07\, -0.12 i & -0.45-0.58 i & -0.089-0.86 i \\
 -0.12-0.074 i & -0.58+0.45i & -0.86+0.089 i \\
 -1.96+2.96 i & 26.81\, +0.97 i & 26.9\, +22.72 i \\
 -2.9-1.96 i & -0.97+26.8 i & -22.7+26.9 i \\
 5.92\, +3.92 i & 1.96\, -53.5 i & 45.4\, -53.8 i \\
 3.92\, -5.92 i & -53.5-1.96 i & -53.8-45.4 i \\
\end{array}
\right) $\\
\hline
        II &  $10^{-3} \left(
\begin{array}{ccc}
 0.012\, -0.026 i & -0.52+0.58 i & -0.87+0.16 i \\
 0.026\, +0.012 i & -0.58-0.52 i & -0.16-0.87 i \\
 2.74\, +5.53 i & -19.7+52. i & -54.5+27.7 i \\
 5.53\, -2.7 i & 52.9\, +19.7 i & 27.7\, +54.5 i \\
 -8.29+4.11 i & -79.4-29.7 i & -41.6-81.8 i \\
 4.1\, +8.29 i & -29.7+79.4 i & -81.8+41.62 i \\
\end{array}
\right)$\\
        \hline
        III & $10^{-3}\left(
\begin{array}{ccc}
 -0.33-1.1 i & -0.1-1.68 i & -0.11-1.51 i \\
 -1.10+0.33 i & -1.68+0.1 i & -1.5+0.11 i \\
 0.15\, -0.26 i & -0.02+1.6 i & -0.03-0.02 i \\
 -0.26-0.15 i & 1.6\, +0.02 i & -0.025+0.03 i \\
 -0.72+1.48 i & 0.15\, -2.1 i & 0.17\, -4.82 i \\
 -1.48-0.72 i & 2.17\, +0.15 i & 4.8\, +0.17 i \\
\end{array}
\right)$\\
\hline
IV & $10^{-3} \left(
\begin{array}{ccc}
 0.68\, +0.25 i & -0.02+0.027 i & -0.027-6.19 i \\
 0.25\, -0.68 i & 0.027\, +0.024 i & -6.19+0.027 i \\
 0.34\, -9.61 i & 0.94\, -9.56 i & 1.06\, +5.42 i \\
 9.61\, +0.34 i & 9.56\, +0.94 i & -5.42+1.06 i \\
 8.24\, +0.70 i & -0.069-38.38 i & -0.078-40.06 i \\
 0.70\, -8.24 i & -38.38+0.069 i & -40.06+0.078 i \\
\end{array}
\right) $\\
\hline
    \end{tabular}
    \caption{Yukawa coupling matrices associated with BPs mentioned in table \ref{tab:caseB}.}
    \label{tab:Yukawa}
\end{table}
 It is important to note here that, in all the sub cases of case B the branching ratio obtained here is of very much improved order which is pretty close to the present sensitivity set by MEG \cite{MEG:2016leq} on Br$(\mu \rightarrow e\gamma)$. For complex $\mathbb{R}$ and first choice of RHN mass values one can notice the green points, allowed by NU lepton mixing restricts the LNV scale from $10^{-3} \,-\,10^{-2}$\,GeV. We also notice a lower bound on the lightest neutrino mass for NH to be around $10^{-4}$\,eV. However for real $\mathbb{R}$, a large branching for $(\mu \rightarrow e\gamma)$ demands a smaller $\mu$ scale which is close to the traditional $\mu$ scale of ISS. Also we do not get such lower bound on the lightest neutrino mass for the consideration of real $\mathbb{R}$. From the results of this section it is quite understandable that, the ISS parameter space yielding an order smaller Br($\mu \to e\gamma$) than the present bound can be probed by MEG II \cite{Cattaneo:2017psr} with an improved sensitivity. 
Neutrino-less-double-beta decay (NDBD) is potentially connected to baryon asymmetry. Since the baryon asymmetry criteria has imposed restriction on the range of lightest neutrino mass, it is instructive to study the effective neutrino mass parameter space which influences the half-life of NDBD amplitude \cite{Mohapatra:1998ye}. The effective neutrino mass governing the NDBD~\cite{RevModPhys.59.671} can be computed using the following prescription.
\begin{equation}
m_{\beta\beta}= \sum_i \left| U_{ei}^2 m_i\right|,
\end{equation}
where, $i = 1,\,2,\,3$, $U$ being the lepton mixing matrix. In this framework, the effective mass would not receive any contribution from the Majorana phases as they are assumed to conserve the CP symmetry. In Fig.~\ref{fig:ndbd} we have shown the variation of effective neutrino mass with the lightest neutrino mass. The figure demonstrates that the suggested sensitivity established by nEXO \cite{nEXO:2018ylp} has the capability to explore a certain region of the parameter space for leptogenesis that is associated with the lightest neutrino mass.

The angle of light-heavy mixing serves as a crucial factor that impacts both the generation of the baryon asymmetry of the universe (BAU) and the control of production rates for Right-Handed Neutrinos (RHNs) at colliders. Unlike the Standard Model (SM) particles, these RHNs are not electrically charged and therefore exhibit highly suppressed direct interactions with SM fields. Their production can only occur through mixing with SM neutrinos \cite{Deppisch:2015qwa,Banerjee:2015gca}. We have determined the magnitude of the light-heavy mixing element, denoted as $|V_{\mu N_{1,2}}|^2$, to be approximately $10^{-7}$ to $10^{-4}$ within the parameter space that corresponds to the observed BAU. This range of $|V_{\mu N_{1,2}}|^2$ values can potentially be explored in forthcoming muon collider experiments, as discussed in a recent publication \cite{Li:2023tbx} and its references. Several distinct signatures of heavy RHNs, dependent on the extent of light-heavy mixing, can be found in references \cite{Basso:2009hf,Li:2023tbx,Banerjee:2015gca,Chakraborty:2018khw} in the context of future lepton colliders, and in references \cite{delAguila:2007qnc,Alva:2014gxa,Pascoli:2018heg} with regards to LHC experiments. At this stage we can be a little optimistic about the validity of our results for ISS parameter space for leptogenesis w.r.t. two complementary searches namely the rare leptonic decay $\mu \to e \gamma$ and searches at Muon Collider. 

\section{Conclusion}
We have proposed the possible scenarios which can account for the observed baryon asymmetry of the Universe in a pure ISS framework. This proposal is based on two independent methods, one by minimizing the huge washout and the other by increasing the order of lepton asymmetry. In both of these cases our conclusion on the success of ISS leptogenesis remains the same. We want to emphasize that ISS parameter space alone can yield the observed baryon to photon ratio through leptogenesis. In order to execute the first approach we have considered non-standard thermal history of the Universe, so that the Hubble expansion rate during leptogenesis can well compete with the large decay rate for leptogenesis. In this case we noticed that, the parameters ($n,\, T_r$) representing the non-standard thermal history of the Universe are able to rescue the ISS parameter space (in terms of $\mu$) which offers successful leptogenesis. 
Next, we consider the second approach where we assume degeneracy among the RHN mass states associated with each generation. This assumption helps us avoid potential cancellations between the lepton asymmetries associated with the PD states of each generation. In most of the cases studied here we have obtained testable parameter space of leptogenesis w.r.t. the LFV search e.g. $\mu \to e\gamma$. To the best of our knowledge, these results are novel and have not been reported earlier. We have found a favourable case for the case A, assigned with $n=3$ both for real and complex rotational matrix $\mathbb{R}$. We have also noticed in case B that, an adequate amount of the aforementioned degeneracy is crucial for the final baryon asymmetry to survive. An important research endeavor would involve exploring the unconventional leptogenesis in ISS framework which can be found elsewhere.

\appendix

\section{Finite temperature effects on resonant leptogenesis}\label{sec:FTbaryonass}
{\color{black}We revisit here a generic case of resonant leptogenesis by including finite temperature correction in a type I seesaw framework with $\mathcal{O}(1)$\,TeV RHN masses. We introduce three RHNs and consider the mass matrix of the same as diagonal with entries $\{M_1$,\,$M_2=M_1+\Delta$,\, $M_3\}$.  
We assume quasi degeneracy ($\frac{\Delta}{M_1}\ll 1$) among the first two RHNs and make the third one substantially heavier for simplicity. This allows to check for any possible finite temperature effects on resonant leptogenesis scenario from both $N_1$ and $N_2$ decays. 


A detailed discussion on the influence of thermal effects on the final baryon asymmetry for low and high scale leptogenesis can be found in refs.\ \cite{Giudice:2003jh,Hambye:2016sby,Granelli:2020ysj}. The Flavored CP asymmetry (\cite{BhupalDev:2014pfm,BhupalDev:2014oar}) in presence of thermal effects is given by \cite{Hambye:2016sby,Granelli:2020ysj}, 
\begin{equation}
\label{eq:CPtemp}
\epsilon_{i}^{\alpha}=\sum_{i\not=j}
{\rm sgn}(M_i-M_j)\,
I_{ij,\alpha\alpha}\,\frac{2x^{(0)}\gamma(z)}
{4\frac{\Gamma_{22}}{\Gamma_{jj}}(x^{(0)}+x_T(z))^2+\frac{\Gamma_{jj}}{\Gamma_{22}}\gamma^2(z)}\,,
\end{equation}
%
where,
\begin{equation}
\label{YukawaPiece}
I_{ij,\alpha\alpha}=\frac{{\rm Im}\left[Y^{\dagger}_{i\alpha}Y_{\alpha j}\left(Y^{\dagger}Y\right)_{ij}\right]
+\frac{M_i}{M_j}{\rm Im}\left[Y^{\dagger}_{i\alpha}Y_{\alpha j}\left(Y^{\dagger}Y\right)_{ji}\right]}{\left(Y^{\dagger}Y\right)_{ii}\left(Y^{\dagger}Y\right)_{jj}}\,.
\end{equation}
%

In Eq.\eqref{eq:CPtemp} the quantity 
$x^{(0)}\equiv \Delta M^{(0)} / \Gamma_{22}$, $\Delta M ^{(0)}$ being the $N_2 - N_1$ mass splitting at zero temperature. Thermal corrections to the $N_2 - N_1$ mass splitting, $\Delta M_T$, 
with the total mass splitting given by   
$\Delta M = \Delta M^{(0)} + \Delta M_T$, 
is present in the expression for
$\epsilon_{(i)}^{\alpha}$ and are 
accounted for by the term $x_T(z)$ \cite{Granelli:2020ysj}: 
\begin{equation}
x_T(z)\equiv \frac{\Delta M_T(z)} {\Gamma_{22}}\simeq\frac{\pi}{4z^2}\sqrt{\left(1-\frac{\Gamma_{11}}{\Gamma_{22}}\right)^2+
4\frac{|\Gamma_{12}|^2}{\Gamma_{22}^2}}\,.
\end{equation}
The function $\gamma(z)$ in Eq. \eqref{eq:CPtemp} determines the thermal effects 
to the $N_{1,2}$ self-energy cut \cite{Hambye:2016sby} and can be expressed as 
    $\gamma(z) \equiv \left\langle \frac{p_\mu L^\mu}{p_\nu q^\nu} \right\rangle\,.$ For further details, see ref.\,\cite{Giudice:2003jh}.

The following set of coupled differential equations (taken from ref.\,\cite{Granelli:2020ysj}) have been numerically solved to find the evolution of temperature dependent asymmetry. 
\begin{eqnarray}
\label{BEsRLThN}
    \frac{d N_{N_i}}{dz}  &=& - \left(D_i\right) (N_{N_i}-N_{N_i}^{eq}), \\
\label{BEsRLThL}
    \frac{dN_{\Delta_\alpha}}{dz}&=& \sum_i \left[- \epsilon_i^{\alpha} D_i (N_{N_i}-N_{N_i}^{eq}) - W_i^D p_{i\alpha} N_{\Delta_\alpha}\right],
\end{eqnarray}
In the above $
    p_{i\alpha} = \frac{\left|Y_{\alpha i}\right|^2v^2}{2\tilde{m}_iM_i}$,
with $\tilde{m}_i\equiv(Y^\dagger Y)_{ii}v^2/2M_i$. $p_{i \alpha}$ is the flavour dependent projection probabilities with
 $p_{i \alpha}$, $i=1,2$, 
$\alpha = e,\mu,\tau$. The quantity $W_i^D$ implies the contribution of inverse decay to the final amount of wash out and $D_i=$ denotes the decay term. Importantly, both $D_i$ and $W_i^D$ contain thermal masses of RHNs and final state leptons and Higgs boson (taken from Ref. \cite{Besak:2012qm}). Here, $N_i$ and $N_{\Delta_\alpha}$ are respectively the number density of i$^\text{th}$ RHN and the value of the asymmetry, both scaled by $n_\gamma$. 


\begin{figure*}\label{fig:tempDass}
\begin{center}
\includegraphics[height=5cm,width=6cm]{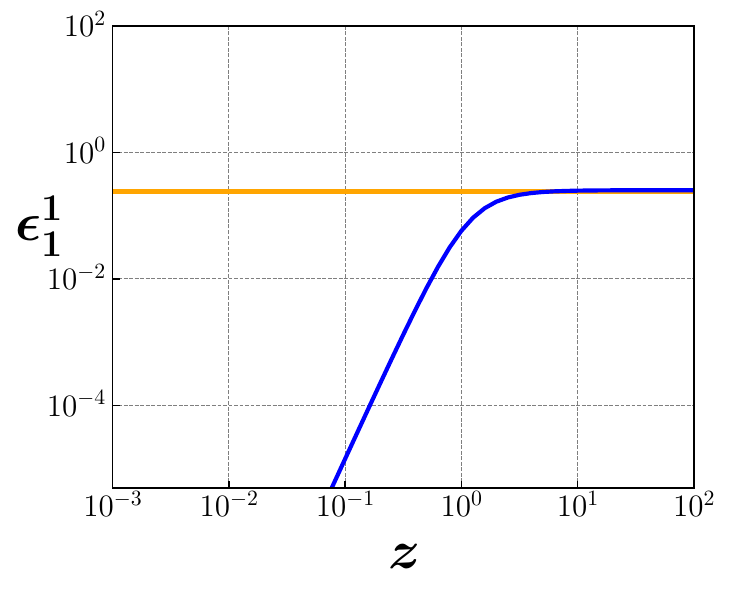}~~~~~~~~~~~~~~~~~~~~
\includegraphics[height=5cm,width=6cm]{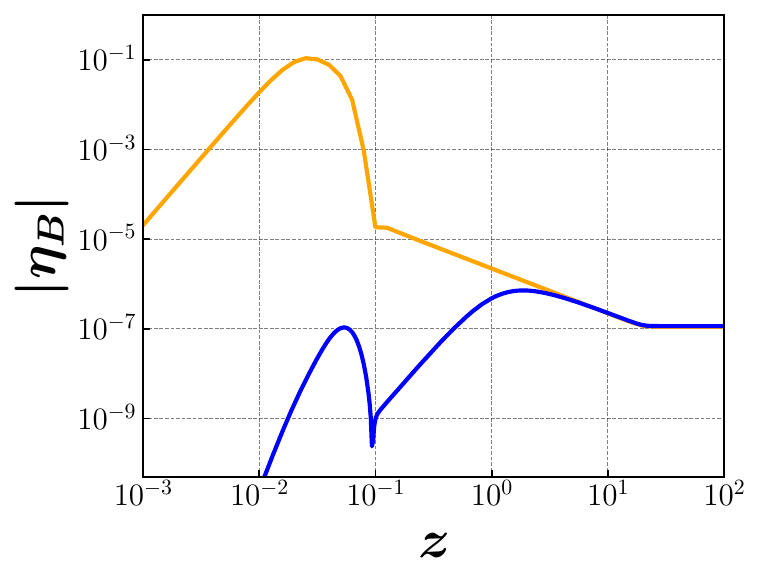}    
\caption{Shows the evolution of CP asymmetry (left) and $\eta_B$ (right) with (blue) and without (orange) temperature correction with respect to ${\rm z}= M/T$.}
\label{fig:finiteT}
\end{center}
\end{figure*}

We present a benchmark scenario to show the impact of thermal effects in the CP asymmetry parameter and final amount of baryon asymmetry. We set,
\begin{align}\label{eq:BPtemp}
   & M_1=1\,\text{TeV},\,\,\Delta M=10^{-7}\,\text{GeV}\nonumber\,\\
   & x=4.18\, +1.68 i,\,\,y=1.95\, +1.68 i,\,\,z=4.0\, +1.45 i\,,
\end{align}
where $x$,\,$y$ and $z$ are the elements of $\mathbb{R}$ in Eq.(\ref{eq:CIRot}).

Next, we compute the CP asymmetry ($\varepsilon_1^e$) associated with the electron flavor from $N_1$ decay and show the same as a function of temperature in the left panel of Fig.\,\ref{fig:tempDass} using Eq.(\ref{eq:CPtemp}). In the relativistic ($z>1$) limit, the CP asymmetry is found to be a decreasing function of temperature while it almost merges with the zero temperature value (Eq.(\ref{eq:asymmetry})) at $z>1$ with very negligible difference not exceeding more than 1$\%$. Similar behaviour of the CP asymmetry with temperature has been also observed in ref.\,\cite{Giudice:2003jh}.

In the right panel of Fig.\,\ref{fig:finiteT}, we estimate the $\eta_B$, incorporating the temperature correction (Eq.(\ref{eq:CPtemp})) and compare it with evolution of the same  without including temperature effect. We utilise the benchmark point as listed in Eq.(\ref{eq:BPtemp}). We notice that the final amount of baryon asymmetry remains more or less the same with or without the inclusion of temperature corrections. Although at high temperature regime ($z\ll 1$), the $\eta_B$ evolution for both the cases seem to be largely different, the final baryon asymmetry is mainly determined only by the later stages of the evolution at relatively small temperatures when the decaying RHN is non-relativistic and thus the corresponding CP asymmetry parameters (with and without temperature corrections) almost matches. Temperature effect might turn crucial for a framework where the RHN mass scale is lower than the EWPT temperature as can be found in \cite{Granelli:2020ysj}. In that case, without temperature correction to the RHN mass, it is not possible to generate lepton asymmetry that can be converted to the observed baryon asymmetry via standard sphaleron process.} 


\bibliography{lepto_inverse.bib}

\end{document}